\documentclass[amsmath,aps,prb,twocolumn,showpacs,superscriptaddress,floatfix]{revtex4}

\usepackage[hypertex]{hyperref}
\usepackage{graphicx}
\usepackage{bm}

\newcommand{\veps}{\ensuremath{\varepsilon}}
\newcommand{\wn}{\ensuremath{~\textrm{cm}^{-1}}}
\newcommand{\te}{\ensuremath{t_{2g}e_g^2}}
\newcommand{\tg}[1][]{\ensuremath{t_{2g}^{#1}}}
\newcommand{\eg}[1][2]{\ensuremath{e_{g}^{#1}}}
\newcommand{\etal}[2][]{\emph{et~al.}#1\cite{#2}}
\renewcommand{\vec}[1]{\ensuremath{\mathbf{#1}}}

\begin{document}
\title{Optical Spectroscopy in CoO: Phonons, Electric, and Magnetic Excitations}
\author{Ch.~Kant}
\author{T.~Rudolf}
\author{F.~Schrettle}
\author{F.~Mayr}
\author{J.~Deisenhofer}
\author{P.~Lunkenheimer}
\affiliation{Experimental Physics~V, Center for Electronic
Correlations and Magnetism, University of Augsburg, 86135~Augsburg,
Germany}
\author{M.~V.~Eremin}
\affiliation{Experimental Physics~V, Center for Electronic
Correlations and Magnetism, University of Augsburg, 86135~Augsburg,
Germany} \affiliation{Kazan State University, 420008 Kazan, Russia}
\author{A.~Loidl}
\affiliation{Experimental Physics~V, Center for Electronic
Correlations and Magnetism, University of Augsburg, 86135~Augsburg,
Germany}

\date{\today}

\begin{abstract}
The reflectivity of single-crystalline CoO has been studied by
optical spectroscopy for wave numbers ranging from 100 to 28,000\wn\
and for temperatures 8 $< T <$ 325~K\@. A splitting of the cubic
IR-active phonon mode on passing the antiferromagnetic phase
transition at $T_N$ = 289~K has been observed. At low temperatures
the splitting amounts to 15.0\wn. In addition, we studied the
splitting of the cubic crystal field ground state of the Co$^{2+}$
ions due to spin-orbit coupling, a tetragonal crystal field, and
exchange interaction. Below $T_N$, magnetic dipole transitions
between the exchange-split levels are identified and the
energy-level scheme can be well described with a spin-orbit coupling
$\lambda = 151.1\wn$, an exchange constant $J = 17.5\wn$, and a
tetragonal crystal-field parameter $D = -47.8\wn$. Already in the
paramagnetic state electric quadrupole transitions between the
spin-orbit split level have been observed. At high frequencies, two
electronic levels of the crystal-field-split $d$-manifold were
identified at 8,000 and 18,500\wn.

\end{abstract}


\maketitle

\section{Introduction}

Strongly correlated transition metal compounds with partly filled
$d$ bands display a variety of properties, interesting for
fundamental research and important for future technological
applications: Colossal magnetoresistance and multiferroicity in the
manganites, high-temperature superconductivity in the cuprates,
exotic superconductivity in the ruthenates and the cobaltites are
illuminating examples. The complexity of the ground state is driven
by strong electronic correlations and a strong interplay between
charge, orbital, spin, and lattice degrees of freedom. In many of
these compounds, orbital degrees of freedom play an essential role
and access to the orbital state can be obtained by studying the
splitting of the $d$ levels, which can reveal the effects of the
crystal-field (CF), spin-orbit coupling (SOC), and exchange-coupling
in the magnetically ordered state. However, experimental methods to
study these local $d$-$d$ excitations are limited. They are electric
dipole forbidden and, hence, very weak when compared to Mott-Hubbard
($d^nd^n$-$d^{n-1}d^{n+1}$) or charge transfer excitations
($d^nd^n$-$d^{n+1}L^-$), which sometimes are in a similar energy
range. Localized $d$-$d$ excitations have  been analyzed using
electron loss spectroscopy\cite{gorschlu94} and, recently, with the
use of nonresonant inelastic X-ray scattering Larson \etal{larson07}
determined the energy scale of $d$-$d$ excitations within the charge
transfer gaps of NiO and CoO\@. The latter results have been
quantitatively explained using a local many body approach by
Haverkort \etal{haverkor07} Note that the late transition metal
monoxides are prototypical correlated electron systems and benchmark
materials for charge-transfer insulators.\cite{zaanen85}

Moreover, transition metal monoxides are regarded as model systems
for spin-phonon coupling effects. The idea of a purely
magnetic-order-induced phonon splitting has been put forth by
Massidda \etal{massidda99} for the antiferromagnetic (AFM)
transition metal monoxides and has been further substantiated in a
recent work by Luo \etal{luo07} The splitting of the transverse
optic modes below the N\'eel temperature $T_N$ has indeed been
experimentally documented by Chung \etal{chung03} in MnO and NiO by
inelastic neutron scattering and by Rudolf \etal{rudolf08} in MnO by
infrared spectroscopy. The splitting of phonon modes has also been
observed in a number of spinel compounds at the onset of AFM
order,\cite{sushkov05,rudolf07,aguilar08} and has been interpreted
in terms of a spin-driven Jahn-Teller (JT)
effect.\cite{yamashit00,tchernys02}

The purpose of this study is to reinvestigate the optical properties
of CoO by infrared (IR) spectroscopy with regard to spin-phonon
coupling effects and $d$-$d$ excitations. CoO has first been
synthesized by Klemm and Sch\"uth.\cite{klemm33} Magnetic
susceptibility
measurements\cite{lablanch51,singer56,mcguire62,uchida64,silinsky81}
show an antiferromagnetic transition at approximately 290~K with a
negative Curie-Weiss temperature of the order of the AFM transition
temperature. The onset of magnetic order is accompanied by a
structural phase transition with a small tetragonal
distortion.\cite{tombs50} At room temperature CoO is paramagnetic
and exhibits the cubic NaCl structure (space group Fm$\bar{3}$m; $a
= 0.42495~\mathrm{nm}$) while at 92~K the lattice parameters of the
tetragonal cell are $a = 0.42552~\mathrm{nm}$ and $c/a = 0.9884$.
Later on, low temperature X-ray diffraction experiments revealed a
rhombohedral distortion in addition to the tetragonal
distortion,\cite{saito66} while high-resolution synchrotron powder
diffraction even manifested a monoclinic symmetry (space group C2/m)
of antiferromagnetic CoO\@.\cite{jauch01} The spin structure of CoO
in the AFM state has been determined by neutron scattering
diffraction by Shull \etal{shull51} On the basis of these data it
has been concluded that the magnetic moments in CoO are arranged in
ferromagnetic (111) planes with the preferred spin direction along
[$\bar{1}\bar{1}7$], which is intermediate to the (111) plane and
the tetragonal axis.\cite{roth58} However, there has been some
dispute about the true magnetic structure and possible multi spin
configurations occurring in
CoO.\cite{roth58a,laar65,herrmann78,ressouch06} Although previous
absorption measurements reported the observation of $d$-$d$
excitations in the paramagnetic (PM) as well as in the AFM
state,\cite{pratt59,milward65,daniel69,austin70} a comprehensive and
unambiguous description of the splittings and determination of the
relevant interaction parameters is still missing. We are able to
describe the observed splittings in the PM and antiferromagnetically
ordered state very well by taking into account SOC, tetragonal
crystal field, and exchange splitting contributions.

\section{Experimental details and sample characterization}

\begin{figure}[t]
\includegraphics[scale=4]{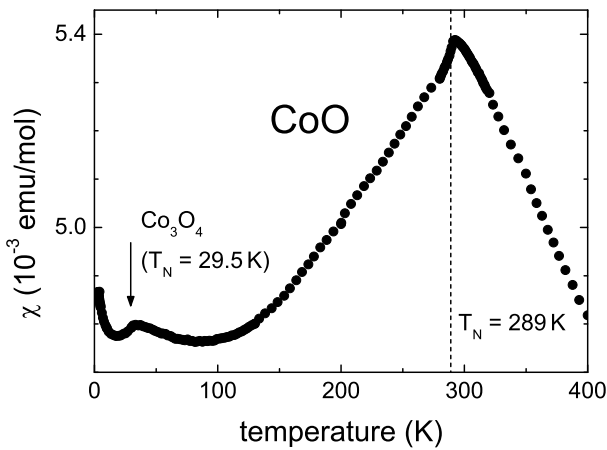}
\caption{\label{fig1}Temperature dependent susceptibility of a
single crystalline CoO platelet, with the external magnetic field
directed perpendicular to the (111) plane. The measurement was
performed at an external magnetic field $\mu_0H$ = 0.1~T\@. The
magnetic ordering temperature of the impurity phase Co$_3$O$_4$ is
indicated by an arrow.}
\end{figure}

High quality single crystals with optical quality (space group
Fm$\bar{3}$m, $a = 0.425~\mathrm{nm}$ at room temperature) in the
form of platelets with dimensions of approximately 1~cm$^2$ and 1~mm
thickness were purchased from MaTecK GmbH\@. Impurities like Fe or
Ni were less than 0.01\%. For characterization, the magnetic
properties were studied using a commercial SQUID magnetometer
(Quantum Design MPMS-5) with external magnetic fields up to 50~kOe.
The heat capacity was measured in a Quantum Design Physical
Properties Measurement System for temperatures from $ 2 < T <
300~\mathrm{K}$. The dielectric properties were determined using a
frequency-response analyzer (Novocontrol) at frequencies between
1~Hz and 1.5~MHz.\cite{schneide01} For these measurements
silver-paint contacts were applied to opposite sides of the
platelets. The reflectivity measurements were carried out using the
Bruker Fourier-Transform Spectrometers IFS 113v and IFS 66v/S, which
both are equipped with a He-flow (4 - 600~K) cryostat. Using
different light sources, different beam splitters and different
detectors, we were able to cover the frequency range from 100\wn\ to
28,000\wn. For the analysis of our reflectivity spectra, we derived
the complex dielectric constant or the complex index of refraction
by means of Kramers-Kronig transformation with a constant
extrapolation towards low frequencies and a smooth power law
extrapolation at high wave numbers.

It is known from previous experiments and publications that even
high-quality single crystals of CoO can suffer from the intergrowth
of Co$_3$O$_4$ clusters. It was shown that small impurity-free
single crystals could only be obtained by annealing CoO crystals in
Co vapor (see Ref.~\onlinecite{silinsky81}). Therefore, we
characterized our samples carefully by measuring the magnetic
susceptibility, the specific heat, and the dielectric properties.
Despite the fact that CoO is a thoroughly studied transition-metal
oxide, we found that the information regarding these basic
properties is incomplete: The temperature dependence of the magnetic
susceptibility of CoO has been studied by Singer\cite{singer56}
between 100~K and 800~K\@. The broad temperature range of this
investigation allowed a rather precise determination of the
Curie-Weiss temperature ($-330~\mathrm{K}$) and paramagnetic moment
($5.25~\mu_B$). A careful susceptibility analysis has been reported
in Ref.~\onlinecite{silinsky81}. These authors removed
nonstoichiometries by heating a small single crystal in Co vapor and
also performed measurements under external stress conditions. We are
not aware of any detailed investigation of the heat capacity
specifically down to low temperatures. Around $T_N$, it has been
investigated in Refs.~\onlinecite{salamon70} and
\onlinecite{massot08}, the specific heat over a broader temperature
range, namely from 100 $< T <$ 500~K is documented in
Refs.~\onlinecite{king57}, \onlinecite{watanabe93}, and
\onlinecite{abarra96}. The dielectric permittivity has been
published by Rao and Smakula\cite{rao65} measuring the dielectric
constant and the dielectric loss of CoO for frequencies between
100~Hz and 1~MHz from liquid nitrogen up to room temperature.

\begin{figure}[b]
\includegraphics[scale=4]{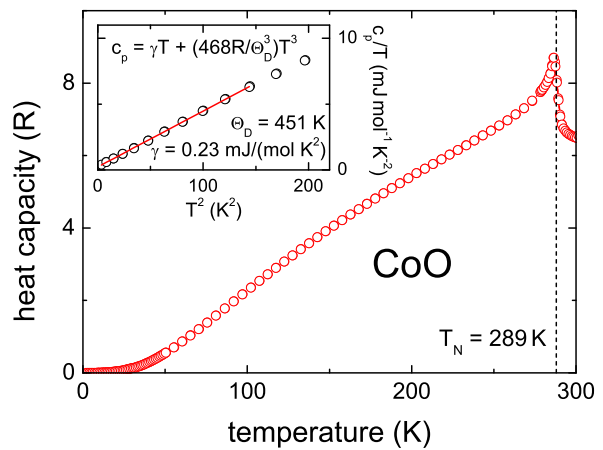}
\caption{\label{fig2}(Color online) Heat capacity of CoO  vs.\
temperature. Inset: plot of $c_{p}/T$ vs.\ $T^2$ (see text). The
best fit of the low temperature specific heat is achieved taking
into account a small linear term $\gamma$ = 0.23~mJ/(mol\,K$^2$)
(solid line in inset).}
\end{figure}

Figure~\ref{fig1} shows the temperature dependence of the magnetic
susceptibility as measured in an external field of $\mu_0H =
0.1~\mathrm{T}$ for temperatures $2.5 < T < 400~\mathrm{K}$, which
is in good agreement with previous publications. The measurement was
performed at an magnetic field $\mu_0 H = 0.1 \mathrm{T}$. The AFM
transition appears as a sharp cusp close to 290~K. Below the
ordering temperature, the susceptibility decreases before a slight
cusp becomes visible close to 30~K, which corresponds to AFM
ordering temperature $T_N$ = 29.5~K of Co$_3$O$_4$.\cite{tristan08}
Towards lower temperatures we observe an increase below about 20~K
which signals a Curie contribution due to free Co spins probably
located in grain boundaries or domain walls. Our susceptibility data
in the range $350 < T < 400~\mathrm{K}$ yields a Curie-Weiss
temperature of approximately $-450~\mathrm{K}$ and an effective
moment $p_\mathrm{eff} = 5.7~\mu_{B}$, which have to be compared to
$\Theta_{CW} = -330~\mathrm{K}$ and $p_\mathrm{eff} = 5.25~\mu_{B}$
obtained by Singer.\cite{singer56} The discrepancy is attributed to
the larger temperature range in the PM phase up to 800~K in the
latter study, which naturally leads to more reliable parameters.
Assuming a spin-only contribution of high-spin Co$^{2+}$-ions with
$S=3/2$ the effective moment $p_\mathrm{eff} = 5.25~\mu_{B}$ results
in an effective $g$-value $g = 2.71$, which indicates a
non-negligible contribution of the orbital momentum.\cite{abragam70}

The temperature dependence of the molar heat capacity $c_p$ is
documented in Fig.~\ref{fig2}. The AFM transition can clearly be
seen. From a closer inspection of the anomaly at the AFM ordering we
determine a phase-transition temperature $T_N = 289~\mathrm{K}$ in
agreement with our susceptibility data. At room temperature, just
above $T_N$, the heat capacity amounts to approximately $6.5~R$ and
is significantly enhanced when compared to a solid with 6 phonon
branches, i.e., $2 \times 3$ degrees of freedom. This enhanced heat
capacity likely results from crystal-field contributions.

At low temperatures the heat capacity can best be fit utilizing a
$T^3$ law and in addition a small linear term. A fit taking into
account only data below 12~K, results in a Debye temperature of
$\Theta_D=451~\mathrm{K}$ (see inset of Fig.~\ref{fig2}), which is
expected from the phonon dynamics. From the normal modes of
vibrations Kushwaha\cite{kushwaha82} calculated $\Theta_{D}(T)$. He
found $\Theta_D \approx 500~\mathrm{K}$ towards 0~K and values
approaching 600~K at the N\'eel temperature, the latter being in
good agreement with published experimental results.\cite{assayag54}
Hence, at low temperatures magnetic contributions seem to play only
a minor role in CoO\@. Obviously, the AFM magnons display a large
gap due to strong SOC, which would explain the absence of any
dispersive magnon contributions at low temperatures and the enhanced
heat capacity at higher temperatures when compared to a non-magnetic
solid. This could also explain the small linear term (see inset in
Fig.~\ref{fig2}). To strengthen these arguments a more detailed
analysis including the heat capacity of phonons, magnons, and
Schottky-like crystal field levels is necessary but beyond the scope
of this paper.

\begin{figure}
\includegraphics[scale=4]{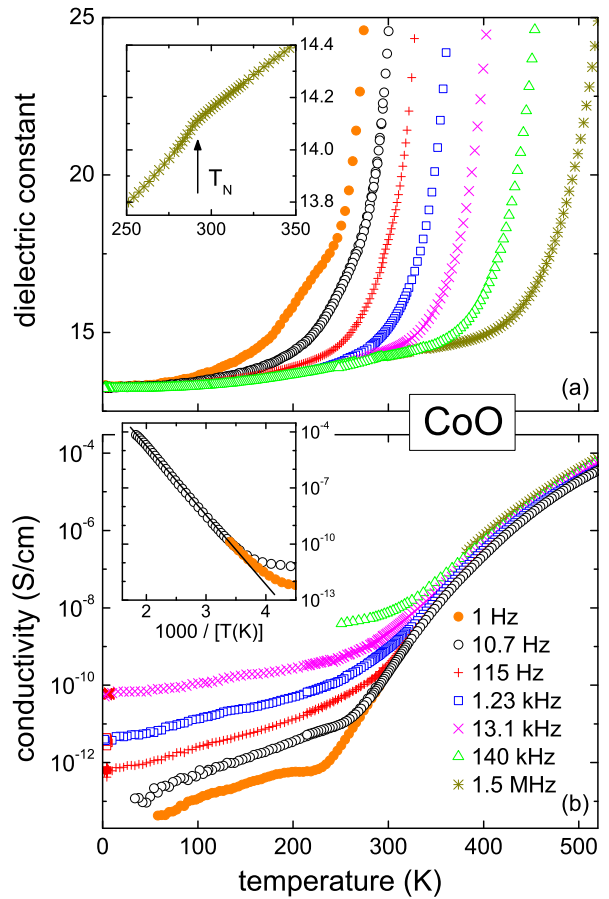}
\caption{\label{fig3}(Color online) Temperature dependence of the
real part of the dielectric constant $\veps'$ [upper frame: (a)] and
real part of the conductivity $\sigma'$ [lower frame: (b)] of CoO
measured at different frequencies between 1~Hz and 1.5~MHz. A (100)
platelet has been investigated with the electrical field
perpendicular to the (100) plane. The inset in (a) demonstrates the
presence of an anomaly in $\veps$'(1.5~MHz)$\approx\veps_{s}$ at the
magnetic phase transition. The inset in (b) shows the temperature
dependence of the $dc$ conductivity in an Arrhenius type
presentation. The line corresponds to a band gap of 1.5~eV.}
\end{figure}

Finally, we also measured the dielectric constant $\veps'$ and the
conductivity $\sigma'$ of CoO between 4 and 500~K and 1~Hz -
1.5~MHz. The results are documented in Fig.~\ref{fig3}. Only below
room temperature and for high measuring frequencies $f$ the static
dielectric constant is approached, yielding approximately $\veps_s$
= 13 at low temperatures [Fig.~\ref{fig3}(a)]. At high temperatures
the real part of the dielectric constant is dominated by \emph{ac}
contributions. It reaches colossal values up to $10^4$ (not shown),
which may be due to hopping conductivity\cite{elliott87} or
Maxwell-Wagner polarization.\cite{lunkenhe02} Only for $f=1.5$~MHz
the room temperature value of the static dielectric constant
$\veps_{s} = 14.2$ can be estimated. This value is slightly enhanced
when compared to published results by Rao and Smakula,\cite{rao65}
who found $\veps_{s} = 12.9$ at room temperature. At relatively high
measuring frequencies, the AFM ordering is displayed by a small cusp
in the temperature-dependent dielectric constant [inset of
Fig.~\ref{fig3}(a)].

The conductivity $\sigma'$ of CoO is shown in Fig.~\ref{fig3}(b).
Here frequency independent $dc$ conductivity dominates above room
temperature, while $ac$ contributions dominate at lower
temperatures. As can be read off close to 100~K in the
Fig.~\ref{fig3}(b), the frequency dependence of the conductivity
deviates from a linear behavior and can best be represented by a
power law, $\sigma \sim \omega^s$, with $s \sim 0.7$, a typical
signature of hopping conduction in disordered
solids.\cite{elliott87} Similar behavior was found in numerous
transition-metal oxides (see, e.g., Ref.~\onlinecite{lunkenhe92}).
The inset of Fig.~\ref{fig3}(b) shows the $dc$ contributions at high
temperatures in an Arrhenius type representation. The conductivity
can well be described assuming a gap of approximately 1.5~eV, which
compares well with published results.\cite{rao65}

\section{Model calculations and analysis}
The reflectivity $R(\omega)$ of an interface is governed by
Fresnel's equations. In case of light impinging on a sample surface
at normal incident, they can be simplified according to
\begin{equation}\label{eqn:nrefl}
  R(\omega)=r(\omega)r^*(\omega)=
  \left|\frac{\sqrt{\veps(\omega)}-\sqrt{\mu(\omega)}}
  {\sqrt{\veps(\omega)}+\sqrt{\mu(\omega)}}\right|^2.
\end{equation}
Here, $r$ is the complex reflectance coefficient, $\veps$ the
dielectric function and $\mu$ the magnetic permeability. In most
compounds, $\mu$ has negligible influence on the reflectivity
spectrum and is usually set to unity in the optical frequency range.
Equation~(\ref{eqn:nrefl}) is valid for all isotropic, homogeneous,
local, and linear materials in the limit of classical
electrodynamics.

In order to analyze our reflectivity data, we use model functions
for both the dielectric function and the magnetic permeability. The
phononic and electronic contributions can be obtained by the
factorized dielectric function\cite{Gervais74}
\begin{equation}\label{eqn:modelEps}
  \veps(\omega)=\veps_\infty \prod_j \frac{\omega^2_{\mathrm{LO}j}-\omega^2-i\gamma_{\mathrm{LO}j}\omega}
  {\omega^2_{\mathrm{TO}j}-\omega^2-i\gamma_{\mathrm{TO}j}\omega},
\end{equation}
where, in case of the phonon modes, $\omega_{\mathrm{TO}j}$,
$\omega_{\mathrm{LO}j}$, $\gamma_{\mathrm{TO}j}$, and
$\gamma_{\mathrm{LO}j}$ can be directly interpreted as
eigenfrequencies ($\omega_j$) and damping constants ($\gamma_j$) of
the transversal (TO) and longitudinal (LO) optical modes,
respectively. $j$ is an index variable which runs over all phononic
and electronic excitations. $\veps_\infty$ arises from
high-frequency electronic absorption processes beyond the phonon
domain. The dielectric strength $\Delta\veps_j$ of excitation $j$
can explicitly be derived from the parameters of the model function
if the resonances are well separated:
\begin{equation}\label{eqn:Dej}
  \Delta\veps_j=\veps_\infty \frac{\omega^2_{\mathrm{LO}j}-\omega^2_{\mathrm{TO}j}}
  {\omega^2_{\mathrm{TO}j}} \prod_{i\geq
  j+1}\frac{\omega^2_{\mathrm{LO}i}}{\omega^2_{\mathrm{TO}i}},
\end{equation}
The effective ionic plasma frequency $\Omega_j$ of each excitation
can then be expressed by
\begin{equation}\label{eqn:wpj}
  \Omega^2_j=\Delta\veps_j\; \omega^2_{\mathrm{TO}j}.
\end{equation}

An alternative way to model the dielectric function is by utilizing
a sum of Lorentz oscillators:
\begin{equation}\label{eqn:modelEpsL}
  \veps(\omega)=\veps_\infty + \sum_j \frac{\Omega_j^2}
  {\omega^2_j-\omega^2-i\gamma_j\omega}
\end{equation}
Here, three parameters are adjustable per mode: $\omega_j$ and
$\gamma_j$ are eigenfrequency and damping of the $j$th resonance,
respectively. The third fit parameter, which enters
Eq.~(\ref{eqn:modelEpsL}), is the plasma frequency $\Omega_j$ of
mode $j$. Comparing the model dielectric functions one can deduce
that Eqs.~(\ref{eqn:modelEps}) and (\ref{eqn:modelEpsL}) become
identical if the damping coefficients of TO and LO modes are equal.

It was shown by Scott\cite{scott71} that the following equation
holds true for the overall plasma frequency in a multimode system:
\begin{equation}\label{eqn:wpcharge}
  \Omega^2=\sum_k \Omega_k^2=\frac{\veps_\infty}{V\veps_\mathrm{vac}}
  \sum_l\frac{(Z_l^*e)^2}{m_l}
\end{equation}
$V$ denotes the unit-cell volume and $Z_l^* e$ the effective charge
of the $l$th ion with mass $m_l$ contributing to a specific phonon
mode. $\veps_\mathrm{vac}$ is the dielectric permittivity of free
space.

Our reflectivity data have been analyzed by utilizing
Eqs.~(\ref{eqn:nrefl}), (\ref{eqn:modelEps}), and
(\ref{eqn:modelEpsL}) and a fit routine which was developed by
Kuzmenko.\cite{kuzmenko}

We also studied phonon eigenfrequencies and damping coefficients as
a function of temperature $T$. To account for purely anharmonic
effects, we assume
\begin{equation}\label{eqn:wanh}
  \omega_{\mathrm{TO}j}(T)=\omega_{0j}\left(
  1-\frac{c_j}{\exp(\Theta_D/T)-1}\right)
\end{equation}
for the temperature dependence of the transverse eigenfrequency
$\omega_{\mathrm{TO}j}$ and
\begin{equation}\label{eqn:ganh}
  \gamma_{\mathrm{TO}j}(T)=\gamma_{0j}\left(
  1+\frac{d_j}{\exp(\Theta_D/T)-1}\right)
\end{equation}
for the temperature dependence of the damping
$\gamma_{\mathrm{TO}j}$ of mode $j$, respectively. $\omega_{0j}$ and
$\gamma_{0j}$ are the eigenfrequency and damping of mode $j$ at
0~K\@.\ $\Theta_D$ denotes the Debye temperature and was treated as
a free fitting parameter as well as the constants $c_j$ and $d_j$,
which determine the strength of the anharmonic contributions.
Detailed calculations of temperature and frequency dependent
anharmonicity can be found in Ref.~\onlinecite{cowley63}.

\section{Results and Discussion}

\begin{figure}
\includegraphics[scale=4]{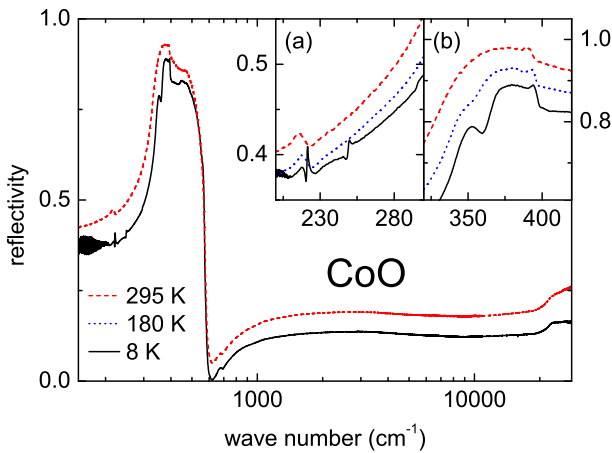}
\caption{\label{fig4}(Color online) Reflectivity of CoO between 150
and 28,000\wn\ at three different temperatures. The curves in the
main frame and inset (b) are successively shifted by an amount of
0.05 for clarity. Inset (a) shows the frequency regime between 200
and 300\wn\ on an expanded scale, while inset (b) provides a closer
look in the frequency range between 320 and 420\wn, where the
splitting of the phonon mode occurs.}
\end{figure}

Figure~\ref{fig4} presents the reflectivity of CoO between 150 and
28,000\wn\ for different temperatures above and below the
antiferromagnetic ordering temperature ($T_N = 289~\mathrm{K}$). At
room temperature the reflectivity is dominated by a broad
reststrahlen band between 300 and 600\wn, which is due to the
optical phonons of the NaCl like structure. At zero wave vector the
two transverse phonons are degenerate and their eigenfrequencies are
determined by the strong increase of the reflectivity close to
300\wn. The frequency of the longitudinal optical phonon is
determined by the steep decrease of the reflectivity at 600\wn.
Ideally, the reflectivity in between these two characteristic
frequencies, where $\veps'$ is negative, should be close to unity,
at least at low temperatures where anharmonic effects are small.
Deviations from this idealized behavior and additional structures
usually result from multiple phonon excitations, which will become
more pronounced with increasing temperature. The observed structure
in the reflectivity on top of the reststrahlen band around 400\wn\
could result from such a two-phonon processes involving zone
boundary optical and acoustical modes, which sum up to a zero
wave-vector excitation with a dipole moment transferred from the
transverse optic phonon mode. Indeed, a week multi-phonon structure
close to 400\wn\ has been calculated by Upadhyay and
Singh.\cite{upadhyay74} However, much stronger peaks in the combined
density of states are calculated to appear close to 435, 500, and
535\wn, which could not be identified unambiguously. In addition, it
has to be stated that the temperature dependence of the observed
structures [inset (b) of Fig.~\ref{fig4}] is not in accord with
anharmonic effects. Multiphonon structures usually become more
significant on increasing temperatures, a fact that definitely is
not observed in CoO\@. The inset (b) of Fig.~\ref{fig4} documents
that the 400\wn\ anomaly remains almost constant from the lowest to
the highest temperatures. Hence, the origin of this structure close
to 400\wn\ remains unsettled. In addition to the reststrahlen band
due to the phonons, an increase in the reflectivity close to
22,000\wn\ signals electronic interband transitions.

\begin{figure}
\includegraphics[scale=4]{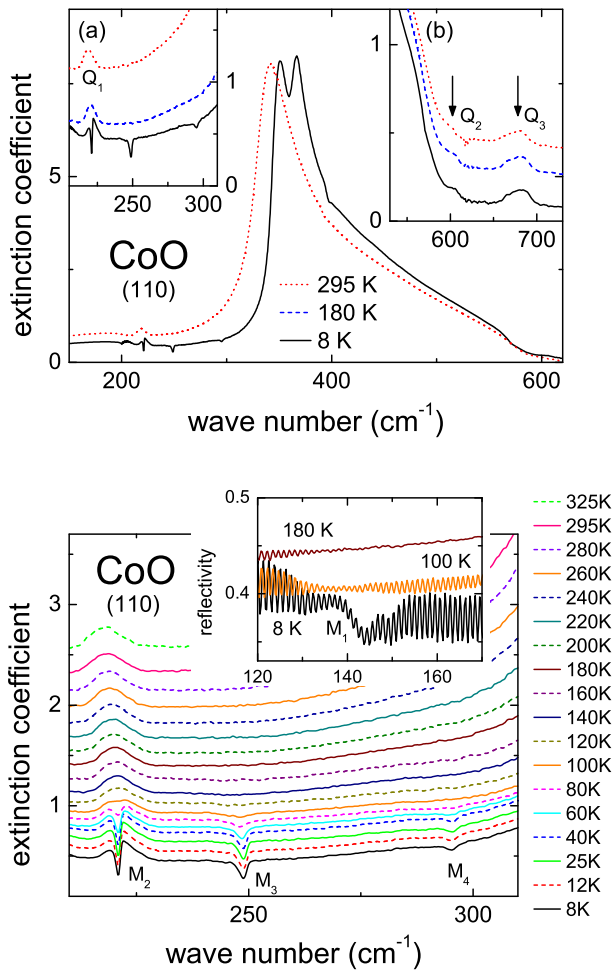}
\caption{\label{fig5}(Color online) Upper panel: Extinction
coefficient $\kappa$ vs.\ wave numbers at 8~K and 295~K\@. The
insets show enlarged regions at (205 - 310\wn) and (530 - 730\wn)
wave numbers for $\kappa$ at 8, 180, and 295~K\@. The curves in both
insets are separated by an amount of 0.2 for clarity. Lower panel:
Sequence of temperature dependent measurements of the extinction
coefficient vs\@. wave numbers between 210 and 320\wn\ and for
temperatures $8 < T < 325~\mathrm{K}$. The data are cumulatively
shifted by an amount of 0.1 for clarity. An electric transition
close to 220\wn\ is visible already at temperatures above the
magnetic phase transition ($T_N = 289~\mathrm{K}$). At 100~K
magnetic dipole transitions evolve, which gain strength on
decreasing temperature. The inset shows an expanded scale of the
reflectivity between 120 and 170\wn\ to demonstrate the appearance
of a further excitation close to 142\wn (see text).}
\end{figure}

At first sight, small additional anomalies close to 220 and 700\wn
can be seen in the frequency-dependent room-temperature reflectivity
displayed in Figure~\ref{fig4}. On cooling and passing the AFM phase
transition no significant shifts or changes are visible, but the
appearance of small additional bands close to 250 and 350\wn\ can be
detected [see insets (a) and (b) of Fig.~\ref{fig4}].
Inset~\ref{fig4}(a) also documents that the rather broad feature of
the 220\wn\ transition is superimposed by a sharp structure when
entering the magnetically ordered phase. The anomaly close to
249\wn\ exhibits a similar shape, namely a dip followed by a peak.
Note that weak features also appear at low temperatures at 142\wn
and 295\wn as shown in Fig.~\ref{fig5}. The interference fringes at
low frequencies are due to the finite thickness of the plan-parallel
sample. At this point we would like to state that the anomalies at
142, 220, 249, 295, and close to 700\wn\ are of electronic origin
and will be discussed later. The additional band close to 350\wn,
which evolves just below the magnetic ordering [see inset (b) of
Fig.~\ref{fig4}], represents the magnetic-order induced phonon
splitting.

We converted the reflectivity spectrum into a frequency dependent
extinction coefficient $\kappa$, which is the imaginary part of the
refractive index, in order to reveal the transition features more
clearly. The reststrahlen band now roughly corresponds to a loss
peak (upper panel of Fig.~\ref{fig5}). It splits at low temperatures
which becomes nicely visible as a clear double peak structure. An
enlarged view of the anomalies between 200 and 300\wn\ is provided
in inset (a). Again, at room temperature ($T > T_N$), only one
single peak is seen, which we denote as $Q_1$ in the following. It
becomes superimposed by an additional negative cusp at low
temperatures. The detailed temperature evolution is shown in the
lower panel of Fig.~\ref{fig5}. In the following, we label the three
negative cusps at 221, 249, and 295\wn\ with $M_2 - M_4$,
respectively. These excitations appear below 100~K only, deep in the
magnetically ordered state. They increase slightly in intensity and
saturate below 25~K, but show no significant shifts in frequency.
The inset in the lower panel documents the occurrence of  a further
excitation ($M_1$) at 142\wn. It behaves similar to $M_2 - M_4$ as
it is dip-like and vanishes at approximately 100~K. $M_1$ strongly
resembles the results of infrared absorption by
Milward,\cite{milward65} who detected a strong absorption at
142.3\wn. Similar findings were revealed in the Raman studies of
Hayes and Perry.\cite{hayes74} This is an experimental evidence,
that $M_1 - M_4$ are intrinsic excitations of CoO, since they appear
at approximately at the same temperature and show no spectral
changes at the ordering temperature of the impurity phase
Co$_3$O$_4$. We further see small kinks at 600 and 680\wn\ [inset
(b) of Fig.~\ref{fig5}], which we will denote as $Q_2$ and $Q_3$,
respectively. They reveal almost no temperature dependence and do
not coincide with calculated multi-phonon bands.\cite{upadhyay74}

In the following we will first discuss the phonon excitations,
before we will turn to the nature of the electronic excitations
occurring below and above the Neel temperature.

\subsection{Phonon excitations}

\begin{table}[b]
\caption{\label{tab1} Phonon excitations in CoO observed in the
present work, compared to reports in literature. All
eigenfrequencies are given in \wn.}
\begin{ruledtabular}
\begin{tabular}{cccccc}
\multicolumn{3}{c}{FIR}&&\multicolumn{2}{c}{neutron
scattering}\\
\multicolumn{2}{c}{(this
work)}&Ref.~\onlinecite{gielisse65}&&\multicolumn{2}{c}{Ref.~\onlinecite{sakurai68}}\\
\cline{1-3}\cline{5-6} 8~K&295~K&295~K&&110~K&425~K
\\\hline

348.1 (TO1)&335.7 (TO)&348.5 (TO)&&348 (TO)&330 (TO)\\
363.1 (TO2)\\
&562.1 (LO)&545.5 (LO)&&524 (LO)\\

\end{tabular}
\end{ruledtabular}
\end{table}

\begin{figure}[b]
\includegraphics[scale=4]{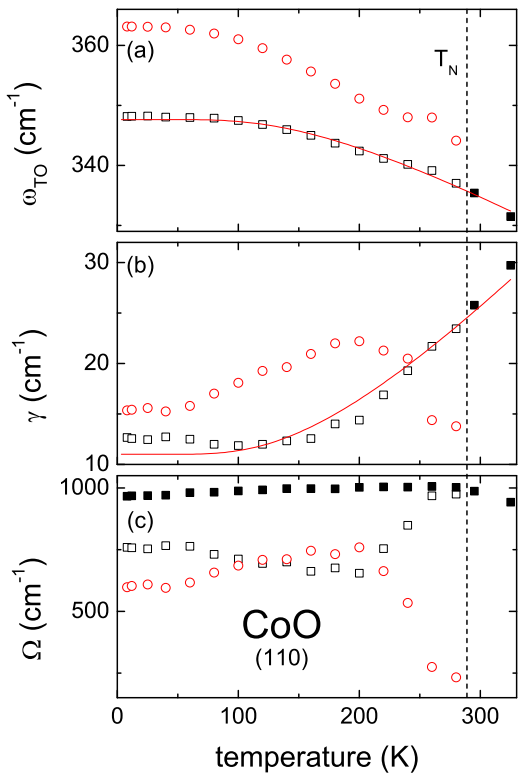} \caption{\label{fig6}(Color
online) Temperature dependencies of eigenfrequency (a), damping
constants (b), and effective ionic plasma frequencies (c) of the
transverse optical phonon modes of CoO\@. The main mode (empty black
squares) and the split-off mode, which appears in the
antiferromagnetic phase (empty red circles), are shown. All
measurements have been performed with single crystalline platelets
with the (110) surface close to normal incidence. The solid lines in
(a) and (b) were calculated assuming a simple anharmonic model [see
Eqs.~(\ref{eqn:wanh}) and (\ref{eqn:ganh})]. Black solid squares
denote the respective quantity above $T_N$ or in (c) the overall
plasma frequency.}
\end{figure}

We tried to fit the reflectivity spectra using a four-parameter fit,
but the anomalies on top of the reststrahlen band together with the
splitting of the phonon modes below $T_N$ did not allow for a
satisfactory fitting procedure. Therefore, we transformed the
reflectivity into dielectric function and fitted these spectra using
a Lorentz fit with three parameters as outlined in
Eq.~(\ref{eqn:modelEpsL}). For comparison, we performed
four-parameter fits in the paramagnetic phase and found good
agreement between the two procedures. The results of the Lorentz
fits are plotted in Fig.~\ref{fig6}. The upper frame (a) gives the
temperature dependence of the TO eigenfrequencies. Assuming that the
main distortion occurring below $T_N$ is tetragonal,\cite{tombs50}
the triply degenerate $T_{1u}$ of the cubic rocksalt structure at
room temperature should split into a doublet and a singlet. Such a
scenario is in agreement with the observed splitting. The mode with
the larger spectral weight at high temperatures smoothly evolves
from the cubic room temperature phase and increases from 335.4\wn\
at room temperature to 348.1\wn\ at liquid He temperatures. Just
below $T_N$ a second mode splits off with a higher eigenfrequency
and increases up to 363.1\wn\ at the lowest temperature. The
splitting of the TO modes in CoO hence amounts 15.0\wn. We would
like to recall, that the low-temperature magnetic phase reveals a
monoclinic symmetry\cite{jauch02} and one certainly can expect a
large number of IR active phonons. However, these additional
splittings are probably below the experimental resolution ($< \sim
0.5\wn$). In comparison to MnO, where the overall splitting of the
TO mode in the antiferromagnetic state was estimated to be
approximately 30\wn,\cite{rudolf08} the splitting is reduced by a
factor of two. Since $\Theta_{CW}/T_N \sim 1$ in CoO, one may infer
that spin frustration indeed may account for the large
magnetic-order induced phonon splitting in MnO.

The temperature dependence of the damping of the transverse phonon
modes is shwon in Fig.~\ref{fig6}(b). The damping of the cubic
$T_{1u}$ continuously decreases towards low temperatures and evolves
again smoothly into the mode with larger spectral weight below $T_N$
and reaches approximately 12.5\wn\ at low temperatures. Such a
behavior is expected for an anharmonic solid and both the
temperature dependence of the eigenfrequency and the damping are
well described by fits [solid lines in Figs.~\ref{fig6}(a) and (b)]
according to Eqs.~(\ref{eqn:wanh}) and (\ref{eqn:ganh}),
respectively. The fit yields a Debye temperature of $\Theta_D$ =
526~K, an enhanced value compared to the low-temperature
thermodynamic Debye temperature (see Fig.~\ref{fig2}) but comparable
to the experimentally determined Debye temperatures at ambient
conditions.\cite{assayag54} The damping of the second mode below
$T_N$ cannot be described by a simple anharmonic behavior.

The temperature dependence of the ionic plasma frequency of the
optical modes is shown in Fig.~\ref{fig6}(c). Both modes roughly are
of equal strength with a value of 700\wn from lowest temperatures up
to 200~K. On further approaching the phase boundary, the main mode
gains weight reaching a value of 990\wn, whereas the split-off mode
rapidly gets suppressed. Due to the overlap of the two modes and the
decreasing weight of the second mode the values above 200~K contain
a larger uncertainty.

Finally, the eigenfrequencies and damping constants of the TO and LO
modes have also been obtained from a four-parameter fit at room
temperature, where only the cubic $T_{1u}$ mode is observable. The
TO and LO eigenfrequencies were determined as 335.7 and 562.1\wn,
respectively. The former value coincides with the one obtained from
the Lorentz fit as expected. In Table~\ref{tab1} these values are
compared to the ones reported earlier by Gielisse \etal{gielisse65}
and Sakurai \etal{sakurai68}. The overall agreement seems
satisfactory, even though the eigenfrequency of the LO mode of the
neutron scattering study is off by almost 8\% when compared to our
result. However, no longitudinal optical eigenfrequency was reported
at room temperature. It should be mentioned that based on the
experimental phonon excitations the phonon dynamics of CoO has been
calculated using lattice-dynamic models of different
complexity\cite{upadhyay74,gupta77,kushwaha82} reaching satisfactory
agreement between experimental data and model calculations. In
Ref.~\onlinecite{upadhyay74} also a detailed calculation of the
two-phonon density of states has been provided. Finally, the lattice
dynamics of CoO has recently been calculated from first
principles.\cite{wdowik07}

The four-parameter fit at room temperature also yielded the
parameters $\veps_{s}$ = 14.0 and $\veps_\infty$ = 5.0, where the
high-frequency dielectric constant has been deduced from fits of the
reflectivity up to 2,000\wn\ (see, e.g., Fig.~\ref{fig4}). The
static dielectric constant compares well to our dielectric result,
$\veps_s$ = 14.2, or to literature, where a value of 12.9
(Ref.~\onlinecite{rao65}) is reported. It is important to note, that
in the four-parameter fit the Lyddane-Sachs Teller relation is
automatically fulfilled and the static dielectric constant follows
from the dielectric strength of the observed phonon mode. It seems
that specifically the dielectric strength and concomitantly the
longitudinal optical phonon frequency is at odds with the values
reported in literature.\cite{gielisse65} From the dielectric
strength the ionic plasma frequency can be directly calculated.
Assuming ideal ionic bonding with valences of $Z = \pm 2$ we expect
an ionic plasma frequency of 2040\wn\ [see
Eq.~(\ref{eqn:wpcharge})]. This has to be compared with the
experimentally observed room temperature value of the ionic plasma
frequency which amounts, $\Omega = 987\wn$. This value is very close
to the one observed in MnO, where $\Omega$ = 1077\wn\ has been
determined.\cite{rudolf08} It indicates strong covalent
contributions to the bonding in CoO\@. Specifically, the effective
valence is found to be $Z^* = 1.0$, much lower than the ideal ionic
valence  of $Z = 2$. From $\gamma$-ray diffraction it has been
concluded that the Co-O interaction is purely ionic.\cite{jauch02}

\subsection{Electronic excitations}

\begin{table*}
\caption{\label{tab2} Electric quadrupole ($Q_i$) and magnetic
dipole ($M_i$) excitations in CoO observed in the present work.
Listed are also electronic resonances and magnon excitations
reported in previous far-infrared absorption, Raman, and neutron
scattering investigations together with the temperatures these
experiments were performed at. All eigenfrequencies are given
in\wn.}
\begin{ruledtabular}
\begin{tabular}{ccccccccc}
\multicolumn{4}{c}{FIR} & \multicolumn{2}{c}{Raman} & \multicolumn{3}{c}{neutron scattering}\\
reflectivity & \multicolumn{3}{c}{transmission}\\
\cline{1-1}\cline{2-4}\cline{5-6}\cline{7-9} (this
work)&Ref.~\onlinecite{milward65}&Ref.~\onlinecite{daniel69}&Ref.~\onlinecite{austin70}&Ref.~\onlinecite{hayes74}&Ref.~\onlinecite{chou76}&Ref.~\onlinecite{sakurai68}&Ref.~\onlinecite{tomiyasu06}&Ref.~\onlinecite{yamani08}\\
8~K&2~K&4.2~K&10~K&20~K&10~K&110~K&10~K&6~K, (1.5\ 1.5\ 0.5)\\
\hline
$\sim$142($M_1$) & 142.3 & & 146 & 143 & 143 & 145-178 & & 163\\
&146.5&&&148\\
220($Q_1$) & & 216 & 215 & & & 214-245 & 218\\
221($M_2$) & & 221 & 221.5 & 221 & 221 & & & 216\\
& & & 233\\
& & & 243\\
250($M_3$) & & 248 & 250 & & & & & 253\\
& & & 260\\
295($M_4$) & & & 296 & 296 & 296 & 340-350 & 313 & 315\\
600($Q_2$) & & & & & 530(?)\\
680($Q_3$)\\
\end{tabular}
\end{ruledtabular}
\end{table*}

\subsubsection{Splittings of the Co$^{2+}$ ground state}

Now, we want to turn to the additional excitation features that are
visible in Figs.~\ref{fig4} and \ref{fig5}. The frequencies derived
from our reflectivity measurements are listed in Table~\ref{tab2}
and compared to excitation energies observed in FIR transmission,
Raman, and neutron scattering experiments. Evidently, many
correspondences between our excitation frequencies and the ones in
literature can be found. Before we discuss in detail the approaches
suggested to describe these excitations, we want to address the fact
that a first distinction between the $Q_1$-$Q_3$ and the $M_1$-$M_4$
excitations can directly be made from the reflectivity data. Looking
at Eq.~\ref{eqn:nrefl} one can see that a contribution to the
magnetic permeability $\mu(\omega)$ like, for example, magnetic
dipole (MD) transitions can lead to a reduction of the reflectivity
and may produce a dip-like feature in the spectrum. MD transition
between $d$ states are symmetry-allowed, but usually are very weak
with an oscillator strength of 10$^{-6}$ and their contribution to
$\mu(\omega)$ does not show up in optical spectra. The situation
changes, however, when AFM order sets in and $\mu(\omega)$ reaches
values which become comparable to $\veps(\omega)$. The transitions
$M_1$-$M_4$ fit well to such a scenario, showing dip-like features
in the reflectivity and appearing only below a temperature of about
100~K, deeply in the AFM phase. Therefore, we assign these features
to MD transitions. Note that similar observations of MD transition
have been reported by H\"ausler \etal{haeussle82} for the related
compound CoF$_2$, where Co has the same electronic configuration as
in CoO. While $M_1 - M_4$ only appear far below $T_N$, the peak-like
transitions $Q_1 - Q_3$ do not show any significant changes at $T_N$
and are, obviously, not related to $\mu(\omega)$ but contribute to
$\veps(\omega)$. It is clear that on-site electric dipole
transitions between the Co $d$-states are parity forbidden, but can
become allowed when the mixing with phonons breaks the inversion
symmetry. Such a mechanism, however, should result in a
characteristic temperature
dependence\cite{tanabe54,liehr57,pratt59,deisenho08}, which is not
observed. Therefore, we assign the transition $Q_1 - Q_3$ to
electric quadrupole transitions, which are symmetry allowed with an
expected oscillator strength of about $10^{-6}$, the same order of
magnitude as the MD transitions.

Let us now turn to the expected level scheme (see Fig.~\ref{fig9})
for Co$^{2+}$ ions in CoO\@. Following Liehr,\cite{liehr63} who
described the three-electron-hole cubic ligand-field spectrum in
full detail, we can assign the lower two eigenstates, at 220 and
600\wn, to transitions between the $^4$F$_{9/2}$ ground state which
is split by the spin orbit coupling. The energy level at 680\wn,
obviously arises from a transition of the ground state to an excited
one which is derived from the $^4$F$_{7/2}$ state.
Liehr\cite{liehr63} deduced his results taking into account SOC in
the atomic limit and then switching on the crystal field. Usually,
the crystal field splitting is calculated from the atomic limit and
then SOC is introduced. In this case and assuming only the lowest
crystal field components, the crystal-field ground state splits into
three levels only, the two highest levels being degenerate. Having
made these qualitative assignments, we will now substantiate our
conclusions by analyzing the energy level scheme of the high-spin
Co$^{2+}$ ion following the approach by Sakurai \etal{sakurai68} To
describe the energy splittings of the groundstate above $T_N$ we
will take into account only SOC, while in the AFM phase an
additional tetragonal crystal field and the effects of exchange
splitting have be considered. Hence, our starting Hamiltonian for
the ground state is given by
\begin{equation}\label{eq:hamilton}
H=\lambda(\vec{S}\vec{L})+D[3L_z^2-L(L+1)]+ 6J \langle
\bm\alpha\vec{S} \rangle (\bm\alpha\vec{S}),
\end{equation}
where $\lambda$ denotes the SOC constant, $D$ is the tetragonal CF
parameter, and the last term is the nearest neighbor exchange
coupling. We want to recall some features of the superexchange
interaction of Co$^{2+}$ ions via oxygen in Co-O-Co fragments (for
short we are using the hole representation \te). From a general
point of view, one can distinguish the following exchange
parameters: $J_{ee}$ between \eg\ sub-shells, $J_{tt}$ between \tg\
holes, and $J_{te}$ between \eg\ and \tg\ sub-systems (this
statement is based on the internal symmetry of the exchange
Hamiltonian; for details see Ref.~\onlinecite{eremin77}). Note that
the \eg\ sub-shells are half filled and that therefore the effective
orbital momentum vanishes. Hence, the leading term in the
superexchange interaction of the Co ions can be written in the
Heisenberg form in first approximation, i.e.,
$H_{ab}^\mathrm{ex}\approx J (\vec{S}_a\vec{S}_b)$, where
$\vec{S}_a$ and $\vec{S}_b$ are the total spins of the \eg\
sub-shell of the ion at site $a$ and $b$, respectively. The exchange
parameters were constrained to one $J$ since $J_{ee}$ is dominating.
Furthermore, it is possible to introduce an effective orbital
momentum $\mathbf{L}$ with $L=1$ for the Co$^{2+}(^4\Gamma)$ state
as well as an effective total momentum $\mathbf{j=L+S}$ (see
Refs.~\onlinecite{abragam70} and \onlinecite{sakurai68}). Using the
wave functions $|j,m_j\rangle$, listed in
Ref.~\onlinecite{abragam70}, where $m_j$ is the quantum number of
the component of \vec{j} along the axis of quantization, it is
straightforward to deduce the energy spectrum $\epsilon(j,m_j)$ of
the Co$^{2+}(^4\Gamma)$ state in crystal and molecular exchange
fields:
\begin{equation} \label{eq:analytic}
\begin{split}
\epsilon(\frac{1}{2},\pm \frac{1}{2})&=\pm \frac{25}{6}J,\\
\epsilon(\frac{3}{2},\pm \frac{3}{2})&=\frac{3}{2}\lambda
\pm\frac{11}{2}
J-\frac{4}{5}D,\\
\epsilon(\frac{3}{2},\pm \frac{1}{2})&=\frac{3}{2}\lambda
\pm\frac{11}{6}
J+\frac{4}{5}D,\\
\epsilon(\frac{5}{2},\pm \frac{5}{2})&=4\lambda \pm\frac{15}{2}J +D,\\
\epsilon(\frac{5}{2},\pm \frac{3}{2})&=4\lambda \pm\frac{9}{2}J -\frac{1}{5}D,\\
\epsilon(\frac{5}{2},\pm \frac{1}{2})&=4\lambda \pm\frac{3}{2}J
-\frac{4}{5}D
\end{split}
\end{equation}
These analytical expressions are very useful as a starting point for
calculations, however, they have to be corrected, since the axis of
magnetization in CoO is not parallel to the tetragonal one. This
leads to the last term in Eq.~\ref{eq:hamilton}. To date, the values
$\bm\alpha$ are still under debate (see
Ref.~\onlinecite{ressouch06}), however, in most publications
\cite{daniel69,chou76} it is believed that the spins point into the
[$\bar{1}\bar{1}7$] direction. Hence we set
$\bm\alpha=\frac{1}{\sqrt{51}} (-1,-1,7)$. The experimentally
observed magnetic dipole transitions listed in Tab.~\ref{tab2} and
located at 142, 221, 249, and 295\wn\ were fitted using
Eq.~\ref{eq:hamilton} with the SOC parameter $\lambda$, the crystal
field parameter $D$ and the magnetic exchange $J$ as free
parameters. In these calculations we assumed that only transitions
from the ground state $\epsilon(1/2,-1/2)$ were observed and that we
detect all possible transitions with $\Delta m_j=0, \pm1$. The best
fit resulted in a levels scheme $\epsilon(1/2,-1/2) = 0\wn$,
$\epsilon(1/2,+1/2) = 145.7\wn$, $\epsilon(3/2,-3/2) = 221.9\wn$,
$\epsilon(3/2,-1/2) = 248.8\wn$, and $\epsilon(3/2,+1/2) = 294.3\wn$
which is shown in Fig.~\ref{fig9}. This calculated level scheme is
very close to the experimentally observed magnetic excitations and
resulted in parameters $\lambda = 151.1\wn$, $J = 17.5\wn$, and $D =
-47.8\wn$. The transition to the level $\epsilon(3/2,+3/2) =
432.9\wn$ is magnetic dipole forbidden and, therefore, can not be
observed in our experiment.

Having derived the splitting parameters in the AFM state by fitting
the energies of the MD transitions, we have to compare these
parameters to the energies of the transitions $Q_1 - Q_3$ which are
also present above the AFM transitions. We identify these
transitions as electric quadrupole transitions between the
spin-orbit split crystal field states. Note that transitions from
the ground to the excited states with effective moment $j=5/2
\Leftrightarrow \Gamma_7, \Gamma_8$ are allowed only via electric
quadrupole mechanisms. This explains why the optical transitions
near 600 and 680\wn\ are not sensitive to the change of
$\mu(\omega)$ upon the phase transition into the antiferromagnetic
state. As can be seen from Eq.~\ref{eq:analytic} by assuming $J = D
= 0$ the level separation between the ground state and the first
excited state above $T_N$ is given by $3/2 \lambda$. In CoO this
immediately results in a SOC constant $\lambda = 146.7\wn$, in good
agreement with the SOC parameter derived from the MD transition at
low temperatures. To handle the slight discrepancies between
calculation and observed energies, a detailed inspection of the
corrections due to \tg-\eg\ and \tg-\tg\ exchange interactions and
the orthorhombic crystal field is necessary, which is out of the
scope of the present paper.

\subsubsection{Higher-lying CF excitations}

\begin{figure}[b]
\includegraphics[scale=4]{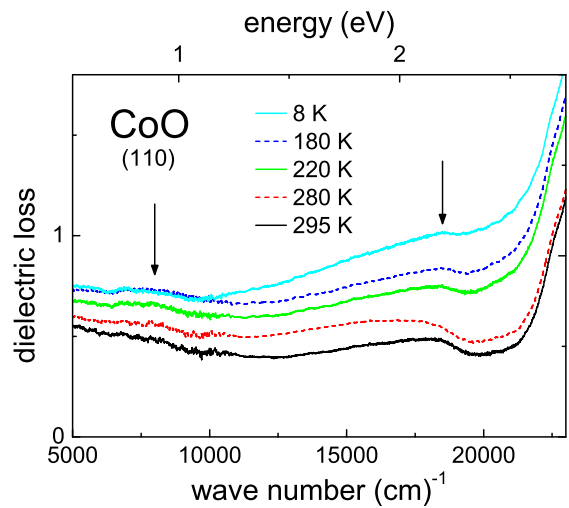}
\caption{\label{fig8}(Color online) Dielectric loss vs\@. wave
numbers for CoO between 5,000 and 23,000\wn. Electronic $d$-$d$
excitations are indicated by arrows.}
\end{figure}

Higher-lying CF excitations have been determined from optical
absorption by Pratt and Coelho.\cite{pratt59} They found two
prominent absorption lines close to 8,000 and 18,500\wn. At low
temperatures the high-energy transition revealed some substructure.
These transitions were identified as transitions from the $^4$F$_1$
ground state to the $^4$F$_2$ and $^4$F$_1$ excited states, which
results in a crystal field parameter of approximately $Dq$ = 900\wn.
Our results are shown in Fig.~\ref{fig8}. In the frequency-dependent
dielectric loss as determined from the reflectivity, these
transitions can only be detected by a close inspection of our
spectra. Slight maxima, which are almost temperature independent and
show no significant changes at the magnetic phase transition, appear
close to 8,000 and 18,500\wn. Beyond 21,000\wn\ ($=
2.6~\mathrm{eV}$) the dielectric loss strongly increases, entering
the frequency regime of strong absorption. The crystal field
excitations are broad and very weak in intensity and do not show any
detectable splittings due to SOC effects. It is clear that these
$d$-$d$ excitations are parity forbidden and gain intensity only via
hybridization with other electronic orbitals or via coupling to
phonons.

It is interesting to note that the optical gap as determined from
the transport measurements rather coincides with the lower band edge
of the 18,500\wn\ transition, which can be located close to
$12,000\wn \sim 1.5~\mathrm{eV}$. It is however unclear, how this
on-site excitation can contribute to the $dc$ hopping transport. The
most plausible explanation of the transport gap certainly is the
formation of an impurity band due to defects.

\section{Summary and concluding remarks}

We performed a careful characterization of single crystalline CoO,
utilizing magnetic and dielectric susceptibility, as well as
thermodynamic measurements. We determined the phase transition into
the AFM state as $T_N = 289~\mathrm{K}$. The low temperature heat
capacity can well be described  by a $T^3$ law with a Debye
temperature $\Theta_D= 451~\mathrm{K}$. It seems that at low
temperatures magnetic excitations play a minor role and do not
contribute to $c_p$. This can be understood from the electronic
excitations, with a series of levels between 150 and 300\wn\ at low
$T$. It documents that SOC is strong compared to the magnetic
exchange. From dielectric spectroscopy we determined $\veps_s =
14.2$ and an electronic gap from the $dc$ transport $E_g =
1.5~\mathrm{eV}$.

We analyzed the phonon dynamics including the spin-phonon coupling
at the antiferromagnetic ordering temperature. Damping and
eigenfrequencies can be described by a normal anharmonic behavior.
However, below $T_N$ a second mode splits off. The splitting amounts
to 15.0\wn\ at low temperatures, considerably lower than in the case
of the isostructural antiferromagnet MnO\@. This behavior might be
explained taking the strong frustration of MnO into account, while
in CoO N\'eel temperature and Curie-Weiss temperature are of the
same order of magnitude. The ionic plasma frequency amounts to about
980\wn, signaling considerable covalent bonding.

\begin{figure}[b]
\includegraphics[scale=0.4]{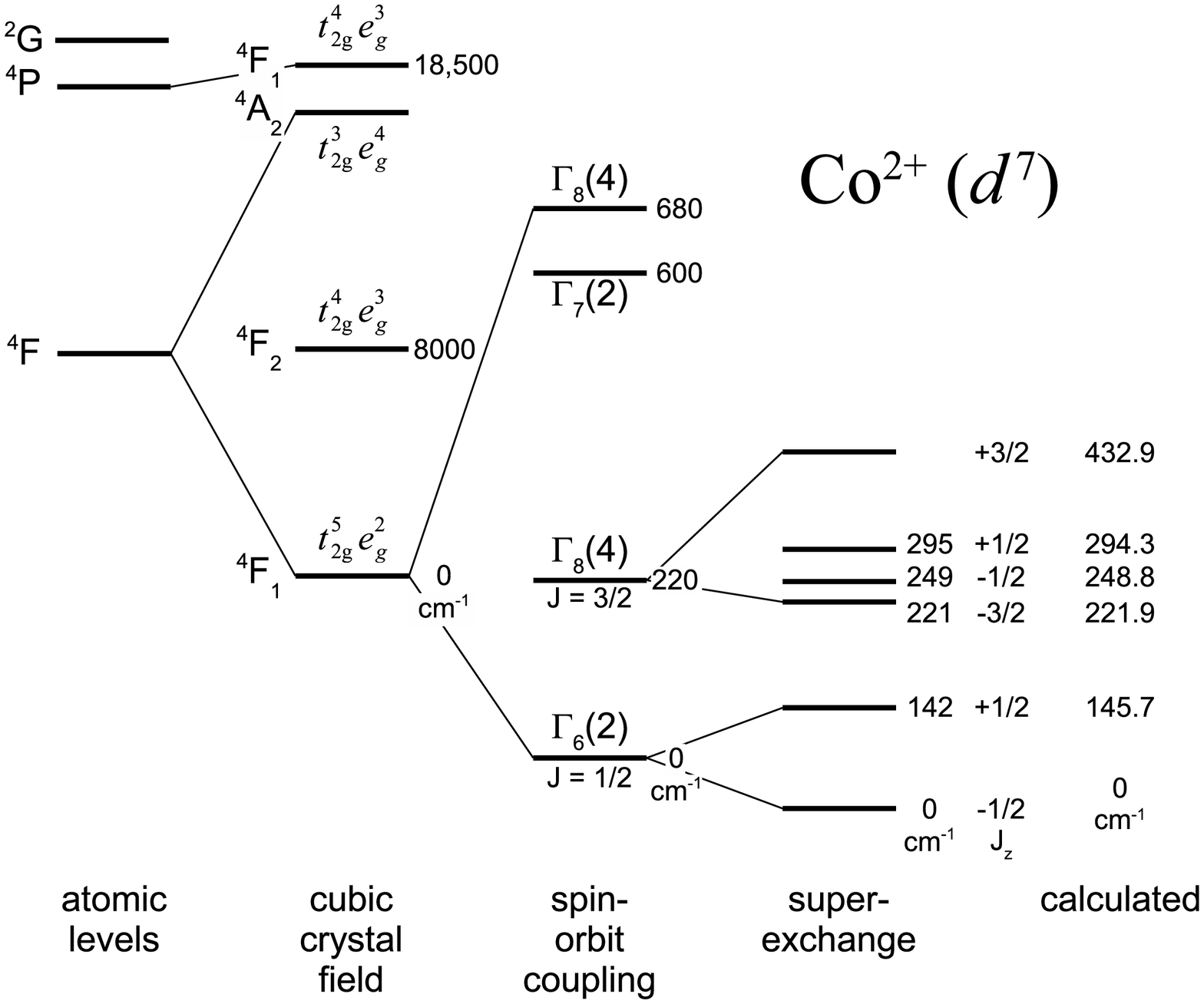}
\caption{\label{fig9}Schematic splitting of the energy levels of the
Co$^{2+}$ ion in CoO\@. The energies are given in \wn\ as determined
in the present work. The magnetic dipole transitions only appear
below 100~K, deep in the magnetically ordered phase. Note that
different energy scales are used to display the effects of different
interactions.}
\end{figure}

In the second part of this work we determined the electronic and
magnetic transitions. The measured reflectivity allows to
discriminate between electric and magnetic dipole transitions. The
complete level scheme of the electronic excitation spectrum of CoO
is plotted in Fig.~\ref{fig9}. Starting from the electronic levels
we determined two crystal field excitations close to 8,000 and
18,500\wn. From these excitations we determine a crystal field
parameter $Dq$ = 900~\wn, in good agreement with values reported
from infrared absorption.\cite{pratt59} These transitions are spin
allowed transitions where only one electron from the \tg\ ground
state (\tg[5]\eg[2]) is excited into an \eg[] level (\tg[4]\eg[3]).
In CoO SOC effects are strong and have to be taken into account. We
determined the splitting of the crystal field ground state by SOC
and identified three excited levels. From the separation of the
first two levels we determine the SOC constant $\lambda =
146.7~\wn$. This value is reasonable compared to the free ion case
where $\lambda_{\mathrm{Ion}} = 176\wn$.\cite{abragam70} Due to the
tetragonal distortion, $\Gamma_7$ and $\Gamma_ 8$ are not degenerate
but are separated by 80\wn. In the free ion case these levels are
degenerate and the separation from the ground state amounts $4
\lambda$. Hence in the atomic limit we would expect one excitation
at 606\wn, instead of two excitations at 600 and 680\wn, which we
identified in CoO, in a crystal with strong covalent bonding.
Finally, we determined a series of magnetic dipole excitations at
142, 221, 249, and 295\wn, which can clearly be identified as
magnetic dipole transitions between the spin-orbit split crystal
field levels whose degeneracy is completely lifted by molecular
exchange fields in the AFM state. The levels for the lowest doublet
and the first excited quartet are also indicted in Fig.~\ref{fig9}.
These transitions appear at low temperatures ($<~100~\mathrm{K}$),
are very weak, and exhibit almost no temperature dependence. We were
able to describe the experimentally observed level scheme
convincingly by a model taking the SOC parameter $\lambda =
151.1\wn$, the magnetic exchange $J = 17.5\wn$, and the tetragonal
crystal field parameter $D = -47.8\wn$ into account. The calculated
energy levels are also displayed in Fig.~\ref{fig9}.

In conclusion, we were able to determine the complete phonon,
electronic and magnetic excitation spectrum of CoO within the
charge-transfer gap by infrared spectroscopy. CoO is a prototypical
and well known example for a strongly correlated system, but its
phononic, electronic, and magnetic excitation schemes offers an
astonishing complexity.

\begin{acknowledgments}
This research has partly been supported by the Deutsche
Forschungsgemeinschaft DFG, via the Collaborative Research Center
SFB 484 (University of Augsburg).
\end{acknowledgments}


\begin{thebibliography}{67}
\expandafter\ifx\csname
natexlab\endcsname\relax\def\natexlab#1{#1}\fi
\expandafter\ifx\csname bibnamefont\endcsname\relax
  \def\bibnamefont#1{#1}\fi
\expandafter\ifx\csname bibfnamefont\endcsname\relax
  \def\bibfnamefont#1{#1}\fi
\expandafter\ifx\csname citenamefont\endcsname\relax
  \def\citenamefont#1{#1}\fi
\expandafter\ifx\csname url\endcsname\relax
  \def\url#1{\texttt{#1}}\fi
\expandafter\ifx\csname urlprefix\endcsname\relax\def\urlprefix{URL
}\fi \providecommand{\bibinfo}[2]{#2}
\providecommand{\eprint}[2][]{\url{#2}}

\bibitem[{\citenamefont{Gorschlueter and Merz}(1994)}]{gorschlu94}
\bibinfo{author}{\bibfnamefont{A.}~\bibnamefont{Gorschluter}}
  \bibnamefont{and} \bibinfo{author}{\bibfnamefont{H.}~\bibnamefont{Merz}},
  \bibinfo{journal}{Phys. Rev. B} \textbf{\bibinfo{volume}{49}},
  \bibinfo{pages}{17293} (\bibinfo{year}{1994}).

\bibitem[{\citenamefont{Larson
  \emph{et~al.}\negmedspace}(2007)\citenamefont{Larson, Ku, Tischler, Lee,
  Restrepo, Eguiluz, Zschack, and Finkelstein}}]{larson07}
\bibinfo{author}{\bibfnamefont{B.~C.} \bibnamefont{Larson}},
  \bibinfo{author}{\bibfnamefont{W.}~\bibnamefont{Ku}},
  \bibinfo{author}{\bibfnamefont{J.~Z.} \bibnamefont{Tischler}},
  \bibinfo{author}{\bibfnamefont{C.-C.} \bibnamefont{Lee}},
  \bibinfo{author}{\bibfnamefont{O.~D.} \bibnamefont{Restrepo}},
  \bibinfo{author}{\bibfnamefont{A.~G.} \bibnamefont{Eguiluz}},
  \bibinfo{author}{\bibfnamefont{P.}~\bibnamefont{Zschack}}, \bibnamefont{and}
  \bibinfo{author}{\bibfnamefont{K.~D.} \bibnamefont{Finkelstein}},
  \bibinfo{journal}{Phys. Rev. Lett.} \textbf{\bibinfo{volume}{99}},
  \bibinfo{pages}{026401} (\bibinfo{year}{2007}).

\bibitem[{\citenamefont{Haverkort
  \emph{et~al.}\negmedspace}(2007)\citenamefont{Haverkort, Tanaka, Tjeng, and
  Sawatzky}}]{haverkor07}
\bibinfo{author}{\bibfnamefont{M.~W.} \bibnamefont{Haverkort}},
  \bibinfo{author}{\bibfnamefont{A.}~\bibnamefont{Tanaka}},
  \bibinfo{author}{\bibfnamefont{L.~H.} \bibnamefont{Tjeng}}, \bibnamefont{and}
  \bibinfo{author}{\bibfnamefont{G.~A.} \bibnamefont{Sawatzky}},
  \bibinfo{journal}{Phys. Rev. Lett.} \textbf{\bibinfo{volume}{99}},
  \bibinfo{pages}{257401} (\bibinfo{year}{2007}).

\bibitem[{\citenamefont{Zaanen
  \emph{et~al.}\negmedspace}(1985)\citenamefont{Zaanen, Sawatzky, and
  Allen}}]{zaanen85}
\bibinfo{author}{\bibfnamefont{J.}~\bibnamefont{Zaanen}},
  \bibinfo{author}{\bibfnamefont{G.~A.} \bibnamefont{Sawatzky}},
  \bibnamefont{and} \bibinfo{author}{\bibfnamefont{J.~W.} \bibnamefont{Allen}},
  \bibinfo{journal}{Phys. Rev. Lett.} \textbf{\bibinfo{volume}{55}},
  \bibinfo{pages}{418} (\bibinfo{year}{1985}).

\bibitem[{\citenamefont{Massidda
  \emph{et~al.}\negmedspace}(1999)\citenamefont{Massidda, Posternak,
  Baldereschi, and Resta}}]{massidda99}
\bibinfo{author}{\bibfnamefont{S.}~\bibnamefont{Massidda}},
  \bibinfo{author}{\bibfnamefont{M.}~\bibnamefont{Posternak}},
  \bibinfo{author}{\bibfnamefont{A.}~\bibnamefont{Baldereschi}},
  \bibnamefont{and} \bibinfo{author}{\bibfnamefont{R.}~\bibnamefont{Resta}},
  \bibinfo{journal}{Phys. Rev. Lett.} \textbf{\bibinfo{volume}{82}},
  \bibinfo{pages}{430} (\bibinfo{year}{1999}).

\bibitem[{\citenamefont{Luo \emph{et~al.}\negmedspace}(2007)\citenamefont{Luo,
  Zhang, and Cohen}}]{luo07}
\bibinfo{author}{\bibfnamefont{W.~D.} \bibnamefont{Luo}},
  \bibinfo{author}{\bibfnamefont{P.~H.} \bibnamefont{Zhang}}, \bibnamefont{and}
  \bibinfo{author}{\bibfnamefont{M.~L.} \bibnamefont{Cohen}},
  \bibinfo{journal}{Solid State Commun.} \textbf{\bibinfo{volume}{142}},
  \bibinfo{pages}{504} (\bibinfo{year}{2007}).

\bibitem[{\citenamefont{Chung
  \emph{et~al.}\negmedspace}(2003)\citenamefont{Chung, Paul, Balakrishnan,
  Lees, Ivanov, and Yethiraj}}]{chung03}
\bibinfo{author}{\bibfnamefont{E.~M.~L.} \bibnamefont{Chung}},
  \bibinfo{author}{\bibfnamefont{D.~M.} \bibnamefont{Paul}},
  \bibinfo{author}{\bibfnamefont{G.}~\bibnamefont{Balakrishnan}},
  \bibinfo{author}{\bibfnamefont{M.~R.} \bibnamefont{Lees}},
  \bibinfo{author}{\bibfnamefont{A.}~\bibnamefont{Ivanov}}, \bibnamefont{and}
  \bibinfo{author}{\bibfnamefont{M.}~\bibnamefont{Yethiraj}},
  \bibinfo{journal}{Phys. Rev. B} \textbf{\bibinfo{volume}{68}},
  \bibinfo{pages}{140406(R)} (\bibinfo{year}{2003}).

\bibitem[{\citenamefont{Rudolf
  \emph{et~al.}\negmedspace}(2008)\citenamefont{Rudolf, {Ch. Kant}, Mayr, and
  Loidl}}]{rudolf08}
\bibinfo{author}{\bibfnamefont{T.}~\bibnamefont{Rudolf}},
  \bibinfo{author}{\bibnamefont{{Ch. Kant}}},
  \bibinfo{author}{\bibfnamefont{F.}~\bibnamefont{Mayr}}, \bibnamefont{and}
  \bibinfo{author}{\bibfnamefont{A.}~\bibnamefont{Loidl}},
  \bibinfo{journal}{Phys. Rev. B} \textbf{\bibinfo{volume}{77}},
  \bibinfo{pages}{024421} (\bibinfo{year}{2008}).

\bibitem[{\citenamefont{Sushkov
  \emph{et~al.}\negmedspace}(2005)\citenamefont{Sushkov, Tchernyshyov,
  Ratcliff~II, Cheong, and Drew}}]{sushkov05}
\bibinfo{author}{\bibfnamefont{A.~B.} \bibnamefont{Sushkov}},
  \bibinfo{author}{\bibfnamefont{O.}~\bibnamefont{Tchernyshyov}},
  \bibinfo{author}{\bibfnamefont{W.}~\bibnamefont{Ratcliff~II}},
  \bibinfo{author}{\bibfnamefont{S.~W.} \bibnamefont{Cheong}},
  \bibnamefont{and} \bibinfo{author}{\bibfnamefont{H.~D.} \bibnamefont{Drew}},
  \bibinfo{journal}{Phys. Rev. Lett.} \textbf{\bibinfo{volume}{94}},
  \bibinfo{pages}{137202} (\bibinfo{year}{2005}).

\bibitem[{\citenamefont{Rudolf
  \emph{et~al.}\negmedspace}(2007{\natexlab{a}})\citenamefont{Rudolf, {Ch.
  Kant}, Mayr, Hemberger, Tsurkan, and Loidl}}]{rudolf07}
\bibinfo{author}{\bibfnamefont{T.}~\bibnamefont{Rudolf}},
  \bibinfo{author}{\bibnamefont{{Ch. Kant}}},
  \bibinfo{author}{\bibfnamefont{F.}~\bibnamefont{Mayr}},
  \bibinfo{author}{\bibfnamefont{J.}~\bibnamefont{Hemberger}},
  \bibinfo{author}{\bibfnamefont{V.}~\bibnamefont{Tsurkan}}, \bibnamefont{and}
  \bibinfo{author}{\bibfnamefont{A.}~\bibnamefont{Loidl}},
  \bibinfo{journal}{Phys. Rev. B} \textbf{\bibinfo{volume}{75}}
  (\bibinfo{year}{2007}{\natexlab{a}});
\bibinfo{author}{\bibfnamefont{J.}~\bibnamefont{Hemberger}},
  \bibinfo{author}{\bibfnamefont{T.}~\bibnamefont{Rudolf}},
  \bibinfo{author}{\bibfnamefont{H.~A.~K.} \bibnamefont{von Nidda}},
  \bibinfo{author}{\bibfnamefont{F.}~\bibnamefont{Mayr}},
  \bibinfo{author}{\bibfnamefont{A.}~\bibnamefont{Pimenov}},
  \bibinfo{author}{\bibfnamefont{V.}~\bibnamefont{Tsurkan}}, \bibnamefont{and}
  \bibinfo{author}{\bibfnamefont{A.}~\bibnamefont{Loidl}},
  \bibinfo{journal}{Phys. Rev. Lett.} \textbf{\bibinfo{volume}{97}}
  (\bibinfo{year}{2006});
\bibinfo{author}{\bibfnamefont{T.}~\bibnamefont{Rudolf}},
  \bibinfo{author}{\bibnamefont{{Ch. Kant}}},
  \bibinfo{author}{\bibfnamefont{F.}~\bibnamefont{Mayr}},
  \bibinfo{author}{\bibfnamefont{J.}~\bibnamefont{Hemberger}},
  \bibinfo{author}{\bibfnamefont{V.}~\bibnamefont{Tsurkan}}, \bibnamefont{and}
  \bibinfo{author}{\bibfnamefont{A.}~\bibnamefont{Loidl}},
  \bibinfo{journal}{New J. Phys.} \textbf{\bibinfo{volume}{9}}
  (\bibinfo{year}{2007}{\natexlab{b}});
\bibinfo{author}{\bibfnamefont{J.}~\bibnamefont{Hemberger}},
  \bibinfo{author}{\bibfnamefont{H.~A.~K.} \bibnamefont{von Nidda}},
  \bibinfo{author}{\bibfnamefont{V.}~\bibnamefont{Tsurkan}}, \bibnamefont{and}
  \bibinfo{author}{\bibfnamefont{A.}~\bibnamefont{Loidl}},
  \bibinfo{journal}{Phys. Rev. Lett.} \textbf{\bibinfo{volume}{98}}
  (\bibinfo{year}{2007}).

\bibitem[{\citenamefont{Aguilar
  \emph{et~al.}\negmedspace}(2008)\citenamefont{Aguilar, Sushkov, Choi, Cheong,
  and Drew}}]{aguilar08}
\bibinfo{author}{\bibfnamefont{R.~V.} \bibnamefont{Aguilar}},
  \bibinfo{author}{\bibfnamefont{A.~B.} \bibnamefont{Sushkov}},
  \bibinfo{author}{\bibfnamefont{Y.~J.} \bibnamefont{Choi}},
  \bibinfo{author}{\bibfnamefont{S.-W.} \bibnamefont{Cheong}},
  \bibnamefont{and} \bibinfo{author}{\bibfnamefont{H.~D.} \bibnamefont{Drew}},
  \bibinfo{journal}{Phys. Rev. B} \textbf{\bibinfo{volume}{77}},
  \bibinfo{pages}{092412} (\bibinfo{year}{2008}).

\bibitem[{\citenamefont{Yamashita and Ueda}(2000)}]{yamashit00}
\bibinfo{author}{\bibfnamefont{Y.}~\bibnamefont{Yamashita}} \bibnamefont{and}
  \bibinfo{author}{\bibfnamefont{K.}~\bibnamefont{Ueda}},
  \bibinfo{journal}{Phys. Rev. Lett.} \textbf{\bibinfo{volume}{85}},
  \bibinfo{pages}{4960} (\bibinfo{year}{2000}).

\bibitem[{\citenamefont{Tchernyshyov
  \emph{et~al.}\negmedspace}(2002{\natexlab{a}})\citenamefont{Tchernyshyov,
  Moessner, and Sondhi}}]{tchernys02}
\bibinfo{author}{\bibfnamefont{O.}~\bibnamefont{Tchernyshyov}},
  \bibinfo{author}{\bibfnamefont{R.}~\bibnamefont{Moessner}}, \bibnamefont{and}
  \bibinfo{author}{\bibfnamefont{S.~L.} \bibnamefont{Sondhi}},
  \bibinfo{journal}{Phys. Rev. Lett.} \textbf{\bibinfo{volume}{88}},
  \bibinfo{pages}{067203} (\bibinfo{year}{2002}{\natexlab{a}});
\bibinfo{author}{\bibfnamefont{O.}~\bibnamefont{Tchernyshyov}},
  \bibinfo{author}{\bibfnamefont{R.}~\bibnamefont{Moessner}}, \bibnamefont{and}
  \bibinfo{author}{\bibfnamefont{S.~L.} \bibnamefont{Sondhi}},
  \bibinfo{journal}{Phys. Rev. B} \textbf{\bibinfo{volume}{66}},
  \bibinfo{pages}{064403} (\bibinfo{year}{2002}{\natexlab{b}}).

\bibitem[{\citenamefont{Klemm and Sch\"uth}(1933)}]{klemm33}
\bibinfo{author}{\bibfnamefont{W.}~\bibnamefont{Klemm}} \bibnamefont{and}
  \bibinfo{author}{\bibfnamefont{W.}~\bibnamefont{Sch\"uth}},
  \bibinfo{journal}{Z. anorg. allg. Chemie} \textbf{\bibinfo{volume}{210}},
  \bibinfo{pages}{33} (\bibinfo{year}{1933}).

\bibitem[{\citenamefont{La~Blanchetais}(1951)}]{lablanch51}
\bibinfo{author}{\bibfnamefont{C.~H.} \bibnamefont{La~Blanchetais}},
  \bibinfo{journal}{J. Phys. Radium} \textbf{\bibinfo{volume}{12}},
  \bibinfo{pages}{765} (\bibinfo{year}{1951}).

\bibitem[{\citenamefont{Singer}(1956)}]{singer56}
\bibinfo{author}{\bibfnamefont{J.~R.} \bibnamefont{Singer}},
  \bibinfo{journal}{Phys. Rev.} \textbf{\bibinfo{volume}{104}},
  \bibinfo{pages}{929} (\bibinfo{year}{1956}).

\bibitem[{\citenamefont{McGuire and Crapo}(1962)}]{mcguire62}
\bibinfo{author}{\bibfnamefont{T.~R.} \bibnamefont{McGuire}} \bibnamefont{and}
  \bibinfo{author}{\bibfnamefont{W.~A.} \bibnamefont{Crapo}},
  \bibinfo{journal}{J. Appl. Phys.} \textbf{\bibinfo{volume}{33}},
  \bibinfo{pages}{1291} (\bibinfo{year}{1962}).

\bibitem[{\citenamefont{Uchida
  \emph{et~al.}\negmedspace}(1964)\citenamefont{Uchida, Nagamiya, Kondoh,
  Nakazumi, Takeda, and Fukuoka}}]{uchida64}
\bibinfo{author}{\bibfnamefont{E.}~\bibnamefont{Uchida}},
  \bibinfo{author}{\bibfnamefont{T.}~\bibnamefont{Nagamiya}},
  \bibinfo{author}{\bibfnamefont{H.}~\bibnamefont{Kondoh}},
  \bibinfo{author}{\bibfnamefont{Y.}~\bibnamefont{Nakazumi}},
  \bibinfo{author}{\bibfnamefont{T.}~\bibnamefont{Takeda}}, \bibnamefont{and}
  \bibinfo{author}{\bibfnamefont{N.}~\bibnamefont{Fukuoka}},
  \bibinfo{journal}{J. Phys. Soc. Jpn.} \textbf{\bibinfo{volume}{19}},
  \bibinfo{pages}{2088} (\bibinfo{year}{1964}).

\bibitem[{\citenamefont{Silinsky and Seehra}(1981)}]{silinsky81}
\bibinfo{author}{\bibfnamefont{P.~S.} \bibnamefont{Silinsky}} \bibnamefont{and}
  \bibinfo{author}{\bibfnamefont{M.~S.} \bibnamefont{Seehra}},
  \bibinfo{journal}{Phys. Rev. B} \textbf{\bibinfo{volume}{24}},
  \bibinfo{pages}{419} (\bibinfo{year}{1981}).

\bibitem[{\citenamefont{Tombs and Rooksby}(1950)}]{tombs50}
\bibinfo{author}{\bibfnamefont{N.~C.} \bibnamefont{Tombs}} \bibnamefont{and}
  \bibinfo{author}{\bibfnamefont{H.~P.} \bibnamefont{Rooksby}},
  \bibinfo{journal}{Nature} \textbf{\bibinfo{volume}{165}},
  \bibinfo{pages}{442} (\bibinfo{year}{1950}).

\bibitem[{\citenamefont{Saito
  \emph{et~al.}\negmedspace}(1966)\citenamefont{Saito, Nakahiga, and
  Shimomur}}]{saito66}
\bibinfo{author}{\bibfnamefont{S.}~\bibnamefont{Saito}},
  \bibinfo{author}{\bibfnamefont{K.}~\bibnamefont{Nakahiga}}, \bibnamefont{and}
  \bibinfo{author}{\bibfnamefont{Y.}~\bibnamefont{Shimomur}},
  \bibinfo{journal}{J. Phys. Soc. Jpn.} \textbf{\bibinfo{volume}{21}},
  \bibinfo{pages}{850} (\bibinfo{year}{1966}).

\bibitem[{\citenamefont{Jauch
  \emph{et~al.}\negmedspace}(2001)\citenamefont{Jauch, Reehuis, Bleif, Kubanek,
  and Pattison}}]{jauch01}
\bibinfo{author}{\bibfnamefont{W.}~\bibnamefont{Jauch}},
  \bibinfo{author}{\bibfnamefont{M.}~\bibnamefont{Reehuis}},
  \bibinfo{author}{\bibfnamefont{H.~J.} \bibnamefont{Bleif}},
  \bibinfo{author}{\bibfnamefont{F.}~\bibnamefont{Kubanek}}, \bibnamefont{and}
  \bibinfo{author}{\bibfnamefont{P.}~\bibnamefont{Pattison}},
  \bibinfo{journal}{Phys. Rev. B} \textbf{\bibinfo{volume}{64}},
  \bibinfo{pages}{052102} (\bibinfo{year}{2001}).

\bibitem[{\citenamefont{Shull
  \emph{et~al.}\negmedspace}(1951)\citenamefont{Shull, Strauser, and
  Wollan}}]{shull51}
\bibinfo{author}{\bibfnamefont{C.~G.} \bibnamefont{Shull}},
  \bibinfo{author}{\bibfnamefont{W.~A.} \bibnamefont{Strauser}},
  \bibnamefont{and} \bibinfo{author}{\bibfnamefont{E.~O.}
  \bibnamefont{Wollan}}, \bibinfo{journal}{Phys. Rev.}
  \textbf{\bibinfo{volume}{83}}, \bibinfo{pages}{333} (\bibinfo{year}{1951}).

\bibitem[{\citenamefont{Roth}(1958{\natexlab{a}})}]{roth58}
\bibinfo{author}{\bibfnamefont{W.~L.} \bibnamefont{Roth}},
  \bibinfo{journal}{Phys. Rev.} \textbf{\bibinfo{volume}{110}},
  \bibinfo{pages}{1333} (\bibinfo{year}{1958}{\natexlab{a}}).

\bibitem[{\citenamefont{Roth}(1958{\natexlab{b}})}]{roth58a}
\bibinfo{author}{\bibfnamefont{W.~L.} \bibnamefont{Roth}},
  \bibinfo{journal}{Phys. Rev.} \textbf{\bibinfo{volume}{111}},
  \bibinfo{pages}{772} (\bibinfo{year}{1958}{\natexlab{b}}).

\bibitem[{\citenamefont{van Laar}(1965)}]{laar65}
\bibinfo{author}{\bibfnamefont{B.}~\bibnamefont{van Laar}},
  \bibinfo{journal}{Phys. Rev.} \textbf{\bibinfo{volume}{138}},
  \bibinfo{pages}{A584} (\bibinfo{year}{1965}).

\bibitem[{\citenamefont{Herrmann-Ronzaud
  \emph{et~al.}\negmedspace}(1978)\citenamefont{Herrmann-Ronzaud, Burlet, and
  Rossat-Mignod}}]{herrmann78}
\bibinfo{author}{\bibfnamefont{D.}~\bibnamefont{Herrmann-Ronzaud}},
  \bibinfo{author}{\bibfnamefont{P.}~\bibnamefont{Burlet}}, \bibnamefont{and}
  \bibinfo{author}{\bibfnamefont{J.}~\bibnamefont{Rossat-Mignod}},
  \bibinfo{journal}{J. Phys. C} \textbf{\bibinfo{volume}{11}},
  \bibinfo{pages}{2123} (\bibinfo{year}{1978}).

\bibitem[{\citenamefont{Ressouche
  \emph{et~al.}\negmedspace}(2006)\citenamefont{Ressouche, Kernavanois,
  Regnault, and Henry}}]{ressouch06}
\bibinfo{author}{\bibfnamefont{E.}~\bibnamefont{Ressouche}},
  \bibinfo{author}{\bibfnamefont{N.}~\bibnamefont{Kernavanois}},
  \bibinfo{author}{\bibfnamefont{L.-P.} \bibnamefont{Regnault}},
  \bibnamefont{and} \bibinfo{author}{\bibfnamefont{J.-Y.} \bibnamefont{Henry}},
  \bibinfo{journal}{Physica B} \textbf{\bibinfo{volume}{385-386}},
  \bibinfo{pages}{394} (\bibinfo{year}{2006}).

\bibitem[{\citenamefont{Pratt and Coelho}(1959)}]{pratt59}
\bibinfo{author}{\bibfnamefont{G.~W.} \bibnamefont{Pratt}} \bibnamefont{and}
  \bibinfo{author}{\bibfnamefont{R.}~\bibnamefont{Coelho}},
  \bibinfo{journal}{Phys. Rev.} \textbf{\bibinfo{volume}{116}},
  \bibinfo{pages}{281} (\bibinfo{year}{1959}).

\bibitem[{\citenamefont{Milward}(1965)}]{milward65}
\bibinfo{author}{\bibfnamefont{R.~C.} \bibnamefont{Milward}},
  \bibinfo{journal}{Phys. Lett.} \textbf{\bibinfo{volume}{16}},
  \bibinfo{pages}{244} (\bibinfo{year}{1965}).

\bibitem[{\citenamefont{Daniel and Cracknel}(1969)}]{daniel69}
\bibinfo{author}{\bibfnamefont{M.~R.} \bibnamefont{Daniel}} \bibnamefont{and}
  \bibinfo{author}{\bibfnamefont{A.~P.} \bibnamefont{Cracknel}},
  \bibinfo{journal}{Phys. Rev.} \textbf{\bibinfo{volume}{177}},
  \bibinfo{pages}{932} (\bibinfo{year}{1969}).

\bibitem[{\citenamefont{Austin and Garbett}(1970)}]{austin70}
\bibinfo{author}{\bibfnamefont{I.~G.} \bibnamefont{Austin}} \bibnamefont{and}
  \bibinfo{author}{\bibfnamefont{E.~S.} \bibnamefont{Garbett}},
  \bibinfo{journal}{J. Phys. C: Solid State Phys.}
  \textbf{\bibinfo{volume}{3}}, \bibinfo{pages}{1605} (\bibinfo{year}{1970}).

\bibitem[{\citenamefont{Schneider
  \emph{et~al.}\negmedspace}(2001)\citenamefont{Schneider, Lunkenheimer,
  Pimenov, Brand, and Loidl}}]{schneide01}
\bibinfo{author}{\bibfnamefont{U.}~\bibnamefont{Schneider}},
  \bibinfo{author}{\bibfnamefont{P.}~\bibnamefont{Lunkenheimer}},
  \bibinfo{author}{\bibfnamefont{A.}~\bibnamefont{Pimenov}},
  \bibinfo{author}{\bibfnamefont{R.}~\bibnamefont{Brand}}, \bibnamefont{and}
  \bibinfo{author}{\bibfnamefont{A.}~\bibnamefont{Loidl}},
  \bibinfo{journal}{Ferroelectrics} \textbf{\bibinfo{volume}{249}},
  \bibinfo{pages}{89} (\bibinfo{year}{2001}).

\bibitem[{\citenamefont{Salamon}(1970)}]{salamon70}
\bibinfo{author}{\bibfnamefont{M.~B.} \bibnamefont{Salamon}},
  \bibinfo{journal}{Phys. Rev. B} \textbf{\bibinfo{volume}{2}},
  \bibinfo{pages}{214} (\bibinfo{year}{1970});
\bibinfo{author}{\bibfnamefont{M.~B.} \bibnamefont{Salamon}},
  \bibinfo{author}{\bibfnamefont{P.~R.} \bibnamefont{Garnier}},
  \bibinfo{author}{\bibfnamefont{B.}~\bibnamefont{Golding}}, \bibnamefont{and}
  \bibinfo{author}{\bibfnamefont{E.}~\bibnamefont{Buehler}},
  \bibinfo{journal}{J. Phys. Chem. Solids} \textbf{\bibinfo{volume}{35}},
  \bibinfo{pages}{851} (\bibinfo{year}{1974}).

\bibitem[{\citenamefont{Massot
  \emph{et~al.}\negmedspace}(2008)\citenamefont{Massot, Oleaga, Salazar,
  Prabhakaran, Martin, Berthet, and Dhalenne}}]{massot08}
\bibinfo{author}{\bibfnamefont{M.}~\bibnamefont{Massot}},
  \bibinfo{author}{\bibfnamefont{A.}~\bibnamefont{Oleaga}},
  \bibinfo{author}{\bibfnamefont{A.}~\bibnamefont{Salazar}},
  \bibinfo{author}{\bibfnamefont{D.}~\bibnamefont{Prabhakaran}},
  \bibinfo{author}{\bibfnamefont{M.}~\bibnamefont{Martin}},
  \bibinfo{author}{\bibfnamefont{P.}~\bibnamefont{Berthet}}, \bibnamefont{and}
  \bibinfo{author}{\bibfnamefont{G.}~\bibnamefont{Dhalenne}},
  \bibinfo{journal}{Phys. Rev. B} \textbf{\bibinfo{volume}{77}},
  \bibinfo{pages}{134438} (\bibinfo{year}{2008}).

\bibitem[{\citenamefont{King}(1957)}]{king57}
\bibinfo{author}{\bibfnamefont{E.~G.} \bibnamefont{King}}, \bibinfo{journal}{J.
  Am. Chem. Soc.} \textbf{\bibinfo{volume}{79}}, \bibinfo{pages}{2399}
  (\bibinfo{year}{1957});
\bibinfo{author}{\bibfnamefont{E.~G.} \bibnamefont{King}} \bibnamefont{and}
  \bibinfo{author}{\bibfnamefont{A.~U.} \bibnamefont{Christensen}},
  \bibinfo{journal}{\emph{ibid.}} \textbf{\bibinfo{volume}{80}},
  \bibinfo{pages}{1800} (\bibinfo{year}{1958}).

\bibitem[{\citenamefont{Watanabe}(1993)}]{watanabe93}
\bibinfo{author}{\bibfnamefont{H.}~\bibnamefont{Watanabe}},
  \bibinfo{journal}{Thermochim. Acta} \textbf{\bibinfo{volume}{218}},
  \bibinfo{pages}{365} (\bibinfo{year}{1993}).

\bibitem[{\citenamefont{Abarra
  \emph{et~al.}\negmedspace}(1996)\citenamefont{Abarra, Takano, Hellman, and
  Berkowitz}}]{abarra96}
\bibinfo{author}{\bibfnamefont{E.~N.} \bibnamefont{Abarra}},
  \bibinfo{author}{\bibfnamefont{K.}~\bibnamefont{Takano}},
  \bibinfo{author}{\bibfnamefont{F.}~\bibnamefont{Hellman}}, \bibnamefont{and}
  \bibinfo{author}{\bibfnamefont{A.~E.} \bibnamefont{Berkowitz}},
  \bibinfo{journal}{Phys. Rev. Lett.} \textbf{\bibinfo{volume}{77}},
  \bibinfo{pages}{3451} (\bibinfo{year}{1996}).

\bibitem[{\citenamefont{Rao and Smakula}(1965)}]{rao65}
\bibinfo{author}{\bibfnamefont{K.~V.} \bibnamefont{Rao}} \bibnamefont{and}
  \bibinfo{author}{\bibfnamefont{A.}~\bibnamefont{Smakula}},
  \bibinfo{journal}{J. Appl. Phys.} \textbf{\bibinfo{volume}{36}},
  \bibinfo{pages}{2031} (\bibinfo{year}{1965}).

\bibitem[{\citenamefont{Tristan
  \emph{et~al.}\negmedspace}(2008)\citenamefont{Tristan, Zestrea, Behr,
  Klingeler, Buchner, von Nidda, Loidl, and Tsurkan}}]{tristan08}
\bibinfo{author}{\bibfnamefont{N.}~\bibnamefont{Tristan}},
  \bibinfo{author}{\bibfnamefont{V.}~\bibnamefont{Zestrea}},
  \bibinfo{author}{\bibfnamefont{G.}~\bibnamefont{Behr}},
  \bibinfo{author}{\bibfnamefont{R.}~\bibnamefont{Klingeler}},
  \bibinfo{author}{\bibfnamefont{B.}~\bibnamefont{Buchner}},
  \bibinfo{author}{\bibfnamefont{H.~A.~K.} \bibnamefont{von Nidda}},
  \bibinfo{author}{\bibfnamefont{A.}~\bibnamefont{Loidl}}, \bibnamefont{and}
  \bibinfo{author}{\bibfnamefont{V.}~\bibnamefont{Tsurkan}},
  \bibinfo{journal}{Phys. Rev. B} \textbf{\bibinfo{volume}{77}},
  \bibinfo{pages}{094412} (\bibinfo{year}{2008}).

\bibitem[{\citenamefont{Abragam and Bleaney}(1970)}]{abragam70}
\bibinfo{author}{\bibfnamefont{A.}~\bibnamefont{Abragam}} \bibnamefont{and}
  \bibinfo{author}{\bibfnamefont{B.}~\bibnamefont{Bleaney}},
  \emph{\bibinfo{title}{Electron Paramagnetic Resonance of Transition Ions}}
  (\bibinfo{publisher}{Oxford University Press}, \bibinfo{year}{1970}).

\bibitem[{\citenamefont{Kushwaha}(1982)}]{kushwaha82}
\bibinfo{author}{\bibfnamefont{M.~S.} \bibnamefont{Kushwaha}},
  \bibinfo{journal}{Physica B \& C} \textbf{\bibinfo{volume}{112}},
  \bibinfo{pages}{232} (\bibinfo{year}{1982}).

\bibitem[{\citenamefont{Assayag and Bizette}(1954)}]{assayag54}
\bibinfo{author}{\bibfnamefont{G.}~\bibnamefont{Assayag}} \bibnamefont{and}
  \bibinfo{author}{\bibfnamefont{H.}~\bibnamefont{Bizette}},
  \bibinfo{journal}{Compt. Rend.} \textbf{\bibinfo{volume}{239}},
  \bibinfo{pages}{238} (\bibinfo{year}{1954}).

\bibitem[{\citenamefont{Elliott}(1987)}]{elliott87}
\bibinfo{author}{\bibfnamefont{S.~R.} \bibnamefont{Elliott}},
  \bibinfo{journal}{Adv. Phys.} \textbf{\bibinfo{volume}{36}},
  \bibinfo{pages}{135} (\bibinfo{year}{1987}).

\bibitem[{\citenamefont{Lunkenheimer
  \emph{et~al.}\negmedspace}(2002)\citenamefont{Lunkenheimer, Bobnar, Pronin,
  Ritus, Volkov, and Loidl}}]{lunkenhe02}
\bibinfo{author}{\bibfnamefont{P.}~\bibnamefont{Lunkenheimer}},
  \bibinfo{author}{\bibfnamefont{V.}~\bibnamefont{Bobnar}},
  \bibinfo{author}{\bibfnamefont{A.~V.} \bibnamefont{Pronin}},
  \bibinfo{author}{\bibfnamefont{A.~I.} \bibnamefont{Ritus}},
  \bibinfo{author}{\bibfnamefont{A.~A.} \bibnamefont{Volkov}},
  \bibnamefont{and} \bibinfo{author}{\bibfnamefont{A.}~\bibnamefont{Loidl}},
  \bibinfo{journal}{Phys. Rev. B} \textbf{\bibinfo{volume}{66}},
  \bibinfo{pages}{052105} (\bibinfo{year}{2002}).

\bibitem[{\citenamefont{Lunkenheimer
  \emph{et~al.}\negmedspace}(1992)\citenamefont{Lunkenheimer, Resch, Loidl, and
  Hidaka}}]{lunkenhe92}
\bibinfo{author}{\bibfnamefont{P.}~\bibnamefont{Lunkenheimer}},
  \bibinfo{author}{\bibfnamefont{M.}~\bibnamefont{Resch}},
  \bibinfo{author}{\bibfnamefont{A.}~\bibnamefont{Loidl}}, \bibnamefont{and}
  \bibinfo{author}{\bibfnamefont{Y.}~\bibnamefont{Hidaka}},
  \bibinfo{journal}{Phys. Rev. Lett.} \textbf{\bibinfo{volume}{69}},
  \bibinfo{pages}{498} (\bibinfo{year}{1992});
\bibinfo{author}{\bibfnamefont{P.}~\bibnamefont{Lunkenheimer}},
  \bibinfo{author}{\bibfnamefont{A.}~\bibnamefont{Loidl}},
  \bibinfo{author}{\bibfnamefont{C.~R.} \bibnamefont{Ottermann}},
  \bibnamefont{and} \bibinfo{author}{\bibfnamefont{K.}~\bibnamefont{Bange}},
  \bibinfo{journal}{Phys. Rev. B} \textbf{\bibinfo{volume}{44}},
  \bibinfo{pages}{5927} (\bibinfo{year}{1991});
\bibinfo{author}{\bibfnamefont{A.}~\bibnamefont{Seeger}},
  \bibinfo{author}{\bibfnamefont{P.}~\bibnamefont{Lunkenheimer}},
  \bibinfo{author}{\bibfnamefont{J.}~\bibnamefont{Hemberger}},
  \bibinfo{author}{\bibfnamefont{A.~A.} \bibnamefont{Mukhin}},
  \bibinfo{author}{\bibfnamefont{V.~Y.} \bibnamefont{Ivanov}},
  \bibinfo{author}{\bibfnamefont{A.~M.} \bibnamefont{Balbashov}},
  \bibnamefont{and} \bibinfo{author}{\bibfnamefont{A.}~\bibnamefont{Loidl}},
  \bibinfo{journal}{J. Phys.: Condens. Matter} \textbf{\bibinfo{volume}{11}},
  \bibinfo{pages}{3273} (\bibinfo{year}{1999});
\bibinfo{author}{\bibfnamefont{P.}~\bibnamefont{Lunkenheimer}}
  \bibnamefont{and} \bibinfo{author}{\bibfnamefont{A.}~\bibnamefont{Loidl}},
  \bibinfo{journal}{Phys. Rev. Lett.} \textbf{\bibinfo{volume}{91}},
  \bibinfo{pages}{207601} (\bibinfo{year}{2003}).

\bibitem[{\citenamefont{Gervais and Piriou}(1974)}]{Gervais74}
\bibinfo{author}{\bibfnamefont{F.}~\bibnamefont{Gervais}} \bibnamefont{and}
  \bibinfo{author}{\bibfnamefont{B.}~\bibnamefont{Piriou}},
  \bibinfo{journal}{Phys. Rev. B} \textbf{\bibinfo{volume}{10}},
  \bibinfo{pages}{1642} (\bibinfo{year}{1974}).

\bibitem[{\citenamefont{Scott}(1971)}]{scott71}
\bibinfo{author}{\bibfnamefont{J.~F.} \bibnamefont{Scott}},
  \bibinfo{journal}{Phys. Rev. B} \textbf{\bibinfo{volume}{4}},
  \bibinfo{pages}{1360} (\bibinfo{year}{1971}).

\bibitem[{\citenamefont{Kuzmenko}()}]{kuzmenko}
\bibinfo{author}{\bibfnamefont{A.}~\bibnamefont{Kuzmenko}},
  \emph{\bibinfo{title}{{RefFIT}}}, \bibinfo{note}{{University of Geneva}},
  \urlprefix\url{{http://optics.unige.ch/alexey/reffit.html}}.

\bibitem[{\citenamefont{Cowley}(1963)}]{cowley63}
\bibinfo{author}{\bibfnamefont{R.~A.} \bibnamefont{Cowley}},
  \bibinfo{journal}{Adv. Phys.} \textbf{\bibinfo{volume}{12}},
  \bibinfo{pages}{421} (\bibinfo{year}{1963}).

\bibitem[{\citenamefont{Upadhyay and Singh}(1974)}]{upadhyay74}
\bibinfo{author}{\bibfnamefont{K.~S.} \bibnamefont{Upadhyay}} \bibnamefont{and}
  \bibinfo{author}{\bibfnamefont{R.~K.} \bibnamefont{Singh}},
  \bibinfo{journal}{J. Phys. Chem. Solids} \textbf{\bibinfo{volume}{35}},
  \bibinfo{pages}{1175} (\bibinfo{year}{1974}).

\bibitem[{\citenamefont{Hayes and Perry}(1974)}]{hayes74}
\bibinfo{author}{\bibfnamefont{R.~R.} \bibnamefont{Hayes}} \bibnamefont{and}
  \bibinfo{author}{\bibfnamefont{C.~H.} \bibnamefont{Perry}},
  \bibinfo{journal}{Solid State Commun.} \textbf{\bibinfo{volume}{14}},
  \bibinfo{pages}{173} (\bibinfo{year}{1974}).

\bibitem[{\citenamefont{Gielisse
  \emph{et~al.}\negmedspace}(1965)\citenamefont{Gielisse, Plendl, Mansur,
  Marshall, Mitra, Mykolajewycz, and Smakula}}]{gielisse65}
\bibinfo{author}{\bibfnamefont{P.~J.} \bibnamefont{Gielisse}},
  \bibinfo{author}{\bibfnamefont{J.~N.} \bibnamefont{Plendl}},
  \bibinfo{author}{\bibfnamefont{L.~C.} \bibnamefont{Mansur}},
  \bibinfo{author}{\bibfnamefont{R.}~\bibnamefont{Marshall}},
  \bibinfo{author}{\bibfnamefont{S.~S.} \bibnamefont{Mitra}},
  \bibinfo{author}{\bibfnamefont{R.}~\bibnamefont{Mykolajewycz}},
  \bibnamefont{and} \bibinfo{author}{\bibfnamefont{A.}~\bibnamefont{Smakula}},
  \bibinfo{journal}{J. Appl. Phys.} \textbf{\bibinfo{volume}{36}},
  \bibinfo{pages}{2446} (\bibinfo{year}{1965}).

\bibitem[{\citenamefont{Sakurai
  \emph{et~al.}\negmedspace}(1968)\citenamefont{Sakurai, Buyers, Cowley, and
  Dolling}}]{sakurai68}
\bibinfo{author}{\bibfnamefont{J.}~\bibnamefont{Sakurai}},
  \bibinfo{author}{\bibfnamefont{W.~J.~L.} \bibnamefont{Buyers}},
  \bibinfo{author}{\bibfnamefont{R.~A.} \bibnamefont{Cowley}},
  \bibnamefont{and} \bibinfo{author}{\bibfnamefont{G.}~\bibnamefont{Dolling}},
  \bibinfo{journal}{Phys. Rev.} \textbf{\bibinfo{volume}{167}},
  \bibinfo{pages}{510} (\bibinfo{year}{1968}).

\bibitem[{\citenamefont{Jauch and Reehuis}(2002)}]{jauch02}
\bibinfo{author}{\bibfnamefont{W.}~\bibnamefont{Jauch}} \bibnamefont{and}
  \bibinfo{author}{\bibfnamefont{M.}~\bibnamefont{Reehuis}},
  \bibinfo{journal}{Phys. Rev. B} \textbf{\bibinfo{volume}{65}},
  \bibinfo{pages}{125111} (\bibinfo{year}{2002}).

\bibitem[{\citenamefont{Gupta and Verma}(1977)}]{gupta77}
\bibinfo{author}{\bibfnamefont{B.~R.~K.} \bibnamefont{Gupta}} \bibnamefont{and}
  \bibinfo{author}{\bibfnamefont{M.~P.} \bibnamefont{Verma}},
  \bibinfo{journal}{J. Phys. Chem. Solids} \textbf{\bibinfo{volume}{38}},
  \bibinfo{pages}{929} (\bibinfo{year}{1977}).

\bibitem[{\citenamefont{Wdowik and Parlinski}(2007)}]{wdowik07}
\bibinfo{author}{\bibfnamefont{U.~D.} \bibnamefont{Wdowik}} \bibnamefont{and}
  \bibinfo{author}{\bibfnamefont{K.}~\bibnamefont{Parlinski}},
  \bibinfo{journal}{Phys. Rev. B} \textbf{\bibinfo{volume}{75}},
  \bibinfo{pages}{104306} (\bibinfo{year}{2007}).

\bibitem[{\citenamefont{Chou and Fan}(1976)}]{chou76}
\bibinfo{author}{\bibfnamefont{H.-h.} \bibnamefont{Chou}} \bibnamefont{and}
  \bibinfo{author}{\bibfnamefont{H.~Y.} \bibnamefont{Fan}},
  \bibinfo{journal}{Phys. Rev. B} \textbf{\bibinfo{volume}{13}},
  \bibinfo{pages}{3924} (\bibinfo{year}{1976}).

\bibitem[{\citenamefont{Tomiyasu and Itoh}(2006)}]{tomiyasu06}
\bibinfo{author}{\bibfnamefont{K.}~\bibnamefont{Tomiyasu}} \bibnamefont{and}
  \bibinfo{author}{\bibfnamefont{S.}~\bibnamefont{Itoh}}, \bibinfo{journal}{J.
  Phys. Soc. Jpn.} \textbf{\bibinfo{volume}{75}}, \bibinfo{pages}{084708}
  (\bibinfo{year}{2006}).

\bibitem[{\citenamefont{Yamani
  \emph{et~al.}\negmedspace}(2008)\citenamefont{Yamani, Buyers, Cowley, and
  Prabhakaran}}]{yamani08}
\bibinfo{author}{\bibfnamefont{Z.}~\bibnamefont{Yamani}},
  \bibinfo{author}{\bibfnamefont{W.}~\bibnamefont{Buyers}},
  \bibinfo{author}{\bibfnamefont{R.}~\bibnamefont{Cowley}}, \bibnamefont{and}
  \bibinfo{author}{\bibfnamefont{D.}~\bibnamefont{Prabhakaran}},
  \bibinfo{journal}{Physica B} \textbf{\bibinfo{volume}{403}},
  \bibinfo{pages}{1406} (\bibinfo{year}{2008}).

\bibitem[{\citenamefont{H\"aussler
  \emph{et~al.}\negmedspace}(1982)\citenamefont{H\"aussler, Lehmeyer, and
  Merten}}]{haeussle82}
\bibinfo{author}{\bibfnamefont{K.~M.} \bibnamefont{H\"aussler}},
  \bibinfo{author}{\bibfnamefont{A.}~\bibnamefont{Lehmeyer}}, \bibnamefont{and}
  \bibinfo{author}{\bibfnamefont{L.}~\bibnamefont{Merten}},
  \bibinfo{journal}{Phys. Stat. Sol. B} \textbf{\bibinfo{volume}{111}},
  \bibinfo{pages}{513} (\bibinfo{year}{1982}).

\bibitem[{\citenamefont{Tanabe and Sugano}(1954)}]{tanabe54}
\bibinfo{author}{\bibfnamefont{Y.}~\bibnamefont{Tanabe}} \bibnamefont{and}
  \bibinfo{author}{\bibfnamefont{S.}~\bibnamefont{Sugano}},
  \bibinfo{journal}{J. Phys. Soc. Jpn.} \textbf{\bibinfo{volume}{9}},
  \bibinfo{pages}{753} (\bibinfo{year}{1954}).

\bibitem[{\citenamefont{Liehr and Ballhausen}(1957)}]{liehr57}
\bibinfo{author}{\bibfnamefont{A.~D.} \bibnamefont{Liehr}} \bibnamefont{and}
  \bibinfo{author}{\bibfnamefont{C.~J.} \bibnamefont{Ballhausen}},
  \bibinfo{journal}{Phys. Rev.} \textbf{\bibinfo{volume}{106}},
  \bibinfo{pages}{1161} (\bibinfo{year}{1957}).

\bibitem[{\citenamefont{Deisenhofer
  \emph{et~al.}\negmedspace}()\citenamefont{Deisenhofer, Leonov, Eremin, {Ch.
  Kant}, Ghigna, Mayr, Iglamov, Anisimov, and van~der Marell}}]{deisenho08}
\bibinfo{author}{\bibfnamefont{J.}~\bibnamefont{Deisenhofer}},
  \bibinfo{author}{\bibfnamefont{I.}~\bibnamefont{Leonov}},
  \bibinfo{author}{\bibfnamefont{M.~V.} \bibnamefont{Eremin}},
  \bibinfo{author}{\bibnamefont{{Ch. Kant}}},
  \bibinfo{author}{\bibfnamefont{P.}~\bibnamefont{Ghigna}},
  \bibinfo{author}{\bibfnamefont{F.}~\bibnamefont{Mayr}},
  \bibinfo{author}{\bibfnamefont{V.~V.} \bibnamefont{Iglamov}},
  \bibinfo{author}{\bibfnamefont{V.~I.} \bibnamefont{Anisimov}},
  \bibnamefont{and} \bibinfo{author}{\bibfnamefont{D.}~\bibnamefont{van~der
  Marell}}, \bibinfo{note}{arXiv:0809.0666v1 (unpublished)}.

\bibitem[{\citenamefont{Liehr}(1963)}]{liehr63}
\bibinfo{author}{\bibfnamefont{A.~D.} \bibnamefont{Liehr}},
  \bibinfo{journal}{J. Phys. Chem.} \textbf{\bibinfo{volume}{67}},
  \bibinfo{pages}{1314} (\bibinfo{year}{1963}).

\bibitem[{\citenamefont{Eremin and Kakitin}(1977)}]{eremin77}
\bibinfo{author}{\bibfnamefont{M.~V.} \bibnamefont{Eremin}} \bibnamefont{and}
  \bibinfo{author}{\bibfnamefont{Y.~V.} \bibnamefont{Kakitin}},
  \bibinfo{journal}{Phys. Stat. Sol. B} \textbf{\bibinfo{volume}{82}},
  \bibinfo{pages}{221} (\bibinfo{year}{1977}).

\end{thebibliography}

\end{document}